# Layer-selective Cooper pairing in an alternately stacked transition metal dichalcogenide

Haojie Guo[1,*], Sandra Sajan[1], Irián Sánchez-Ramírez[1], Tarushi Agarwal[2], Alejandro Blanco Peces[1], Chandan Patra[2], Maia G. Vergniory[1,3], Rafael M. Fernandes[4,5], Ravi Prakash Singh[2], Fernando de Juan[1,6], Maria N. Gastiasoro[1,*], Miguel M. Ugeda[1,6,7,*]

[1]*Donostia International Physics Center, Paseo Manuel de Lardizábal 4, 20018 San Sebastián, Spain.*

[2]*Department of Physics, Indian Institute of Science Education and Research Bhopal, 462066 Bhopal, India*

[3]*Département de physique et Institut quantique, Université de Sherbrooke, QC J1K 2R1 Sherbrooke, Canada*

[4] *Department of Physics, The Grainger College of Engineering, University of Illinois Urbana-Champaign, Urbana, Illinois 61801, USA.*

[5]*Anthony J. Leggett Institute for Condensed Matter Theory, The Grainger College of Engineering, University of Illinois Urbana-Champaign, Urbana, Illinois 61801, USA.*

[6]*Ikerbasque, Basque Foundation for Science, 48013 Bilbao, Spain*

[7]*Centro de Física de Materiales, Paseo Manuel de Lardizábal 5, 20018 San Sebastián, Spain*

*\*Corresponding Authors: haojie.guo@dipc.org, maria.ngastiasoro@dipc.org, mmugeda@dipc.org*

*Multigap superconductivity emerges when superconducting gaps form on distinct Fermi surfaces. Arising from locally overlapping atomic orbitals, multiple superconducting bands introduce a new internal degree of freedom in the material that, however, escapes external control due to their coexistence in real space in the known multigap superconductors. Here, we show that the layered superconductor 4Hb-TaSSe — composed of alternating trigonal (H) and octahedral (T) polymorph layers — is a multigap superconductor, featuring two weakly coupled superconducting condensates with distinct properties, spatially separated in alternating layers. Using high-resolution quasiparticle tunneling and Andreev reflection spectroscopy in the two polymorph layers, we identify two superconducting gaps that vary in size and internal structure. The intrinsic Cooper pairing in each polymorph is corroborated by the temperatures and magnetic fields at which the gaps open up, which differ in each polymorph layer and show opposing resilience to these parameters. This behavior enables selective external actuation upon the condensates. Our theoretical model based on ab-initio calculations reproduces key features of the observed superconducting gaps in the presence of finite interlayer hybridization and explains the unusually high critical field observed in the T-layer. Our results establish TMD polymorphs as platforms for engineering tunable multigap superconductors, offering new opportunities in layered superconducting device architectures.*

The presence of several superconducting bands in a material can trigger genuinely novel phenomena such as increased critical fields and robust pairing against disorder[1,2], increased transition temperatures[3,4], sign-changing gap structures[5], or spontaneous breaking of time-reversal symmetry[6]. Indeed, in a multiband system, qualitatively different gap structures with the same symmetry are possible, giving rise to states with no counterpart in single-band superconductors, such as nodal s-wave and nodeless d-wave superconducting states[7–9]. The nature of the interactions driving the Cooper pairing and the orbital character of the different bands at the Fermi surface ultimately determine the superconducting order parameter in multigap systems. In most cases, the electronic bands are formed from the overlap of *local* atomic orbitals, which determine the matrix elements of intra- and interband pairing and scattering, and result in coupled condensates coexisting in real space (albeit separated in momentum space). This results in an internal degree of freedom in multigap superconductors of potential interest for superconducting-based technologies. In the known multigap superconductors such as $MgB_2$, Fe-based and Kagome compounds, the orbitals involved in the superconducting bands are provided by either one atomic species or by neighboring species within the unit cell and, therefore the superconducting condensates coexist in real space. This makes the development of strategies to independently manipulate these condensates challenging.

In this arena, van der Waals materials provide a promising platform for achieving spatial separation of superconducting condensates across different layers. Specifically, certain polytypes of transition metal dichalcogenides (TMDs) are well-suited for this purpose, as they contain transition metals (such as Ta or Nb) situated in distinct chemical environments and contributing to superconducting bands that are mostly localized in separate layers (polymorphs) within the unit cell. The 4Hb polytype is especially promising in this regard as the unit cell is composed by alternating trigonal (H) layers and octahedral (T) layers (Fig. 1a). Furthermore, the 4Hb-$TaS_2$ compound has been reported to host a chiral superconducting state in the H layers that breaks time reversal symmetry[10,11], with signatures of nematicity[12] and topological features[13]. However, no intrinsic superconductivity has been yet reported for the T-type layer in the 4Hb polytype.

Here, we provide experimental evidence for spatially separated and weakly coupled superconducting condensates in the T- and H-polymorph layers of the 4Hb-TaSSe isovalent alloy[14–16]. Using high-resolution scanning tunneling microscopy/spectroscopy (STM/STS) and Andreev reflection spectroscopy (AR), we find marked distinct superconducting features in each polymorph. Our spectroscopic measurements at 340 mK unveil two different superconducting gaps in size and structure in each polymorph; we observe a gap of 0.46 meV with internal structure in the H-type polymorph and a smaller, nodal gap of 0.23 meV in the T-type polymorph. AR spectroscopy measurements on these polymorphs reveal distinct structures of the Andreev states within the superconducting gaps, consistent with the distinct gaps observed in quasiparticle tunneling spectra. Strikingly, under an out-of-plane magnetic field ($H_\perp$), the larger gap in the H-polymorph is suppressed by 1 T, while the smaller gap in the T- polymorph survives up to 2.6 T. This implies intrinsic, robust pairing in the T layer - absent in isoelectronic 4Hb-TaS$_2$ - and suggests different superconducting regimes operating in this material that can be independently triggered with temperature and applied magnetic field. Our *ab initio* calculations reveal that despite finite H-T interlayer hybridization, parts of the Fermi surface retain dominant T-layer character with low Fermi velocity, originating from quasi-isolated Star-of-David orbitals. This can explain the unusually high critical field of the T-layer and establishes 4Hb-TaSSe as a model system for layer-selective Cooper pairing.

## Results

**Electronic structure of the polymorphs**

Our experiments were carried out in high-quality crystals of the isovalent alloy 4Hb-TaSSe grown by chemical vapor transport (CVT) (methods section and supplementary note 1). The 4Hb polytype belongs to the hexagonal P6$_3$/mmc space group and features an alternating stacking of two H-type and two T-type layers along the *c*-axis (Fig. 1a). After mechanical exfoliation of the TaSSe crystals in ultra-high-vacuum (UHV) conditions, the T- and H-TaSSe polymorph layers become exposed (Fig. 1b), enabling their independent structural and electronic characterization by STM/STS. Figures 1c,d show atomically resolved STM images of the H- and T-type polymorphs, respectively. The H-type surface (Fig. 1c) reveals an atomic registry with a heterogeneous corrugation likely due to the S-Se random arrangement in the chalcogen plane, and also influenced by the charge density wave (CDW) of the T-layer underneath (Fig. S2), which is suggestive of finite hybridization between the two polymorph layers. The H-layer in 4Hb-TaSSe, however, shows no trace of intrinsic CDW order, in contrast to the H-layers in this compound's end members 4Hb-TaS$_2$ and 4Hb-TaSe$_2$, where 2×2 and 3×3 CDWs develop, respectively[17,18]. In contrast, the T-TaSSe layer (Fig. 1d) exhibits a prominent CDW with a $(\sqrt{13}\times\sqrt{13})a_0$-13.9° periodicity, commonly referred to as Star-of-David (SoD). Two distinct dark and bright SoD clusters are discernible, which are related to their local electronic structure, as described below.

The large-scale electronic structure of these two types of layers were characterized by STM differential conductance (d$I$/d$V(V)$) measurements. Figures 1e,f show representative d$I$/d$V$ spectra acquired on the H- and T-layer, respectively. Both layers exhibit a metallic character, i.e., a finite density of states (DOS) at the Fermi level ($E_F$), which enables the development of superconductivity in both layers. Unlike the H layer, the electronic structure of the T-layer is spatially dependent along different SoD patterns, dominated by two broad peaks at ± 0.25 V (labeled as $V_1$ and $C_1$) with a finite DOS at $E_F$. A sharp resonance (10 ± 2 meV) is also observed at some of the SoDs. As seen in the d$I$/d$V(x,V)$ plot acquired along a high-symmetry direction of the CDW shown in Fig. 1g, the intensity of $V_1$ and $C_1$ as well as the presence of the zero-bias resonance varies between consecutive SoDs. In particular, only ~58% of the SoDs in this layer show the resonance, as shown in the d$I$/d$V(0)$ map in Fig. 1h. The DOS at $E_F$ is, therefore, mostly concentrated on these SoD centers, similarly to the case of the related 1T polytype of TaSSe[19]. The intensity correlations among the three main peaks observed in the STS is further analyzed via constant-height d$I$/d$V(V)$ maps at the relevant energy position of the peaks (supplementary note 3).

**Superconductivity in the T-layer of 4Hb-TaSSe**

We now focus on the electronic structure of the T- and H- type layers near $E_F$, aiming at locally probing superconductivity of this isovalent alloy at 0.34 K, our STM base temperature. Previous transport studies reported the presence of superconductivity in 4Hb-TaS$_{2-x}$Se$_x$ for a wide $x$-range with an optimal critical temperature ($T_C$) for $x$ = 1, i.e., 4Hb-TaSSe[14,15]. The DOS of the T-layer around $E_F$ shows large spatial variations within the nm-scale due to the presence of sharp resonances in roughly half of the SoDs (Figs. 1f-h), as shown in Fig. 2a. Figure 2b shows two representative d$I$/d$V$ spectra taken consecutively on the SoDs encircled in Fig. 2a. Regardless of the presence of the zero-bias resonance, the DOS at 0.34 K always shows a homogeneous V-shape dip of $\Delta_T$ = 0.23 ± 0.04 mV (peak-to-peak, see supplementary note 4) with a finite conductance at zero energy. The dip feature is typically bound by peaks consistent with the coherence peaks of a superconducting gap (see zoomed spectra in Fig. 2c).

To corroborate the superconducting origin of the V-shaped dip feature, we performed height-dependent STS on the T-layer using a metallic, non-superconducting tip ranging from the tunneling regime (~MΩ) to the contact regime (≲ 10 kΩ). This technique, also known as Andreev reflection (AR) spectroscopy, enables to unambiguously prove the superconducting origin of the gap feature while, simultaneously, providing key information about the structure of the order parameter[20–28]. Figure 2d shows a representative set of d$I$/d$V$ spectra as the tip's regime changes from tunneling (bottom) to contact (top). The development of in-gap conductance triggered by Andreev reflections is far from homogenous as two symmetric peaks located at ± 0.08 mV first emerge in the tunneling regime followed by a third peak centered at $E_F$, with all finally merging in the contact regime. These in-gap features appear consistently, independent on the specific location on the T-phase. The observation of ARs provides unambiguous evidence for the superconducting nature of the *V*-shaped dip in the T-layer. This is further supported by $H_\perp$-dependent AR measurements, which confirm that ARs are suppressed by external magnetic fields (Fig. S5a).

Regarding the structure of the superconducting gap observed in the T-layer, we analyze the quasiparticle spectrum as well as the evolution of the AR spectra shown in Figures 2c,d, respectively. First, we performed a systematic fitting analysis on a statistically significant set of d$I$/d$V$ spectra using a Dynes function[24], comparing both isotropic and nodal order parameters (details in supplementary note 6). The analysis reveals that in 95% of the cases (38/40 curves), the nodal structure quantitatively fits better the superconducting gap of the T-layer. We have tested our fit procedure against the conventional superconductor Al(111), where we find instead only the isotropic gap function is needed to reproduce the experimental gap value and shape (supplementary note 7). The evolution of ARs with the tip-sample distance (Fig. 2d) also departs from the expected behavior for a nodeless *s*-wave superconductor-metal interface. According to the simple 1D Blonder-Tinkham-Klapwijk (BTK) model[23], Andreev states in an isotropic *s*-wave superconductor steadily fill the superconducting gap upon reducing the barrier strength Z, until reaching a "plateau" shape conductivity at $Z = 0$ (Fig. S8a). Instead, nodal or sign-changing gap structures show inhomogenous in-gap features like those observed here (Fig. S8b). Various theoretical frameworks using different approaches have shown, indeed, the described behavior of ARs for those gap structures[20,27–29]. Experimentally, AR spectra recently reported on magic-angle twisted bilayer graphene, a superconductor with a proposed nodal structure, show striking resemblance with a two-peak structure[24]. Although the exact structure of the superconducting gap function will require further experimental and theoretical efforts, our current findings advocate for a superconductor that falls beyond the isotropic, full-gap *s*-wave gap structure. This result is in contrast to the superconductivity in the 1T polytype of the same TaSSe alloy, which shows a conventional s-wave full gap and a homogeneous filling of the Andreev states[19].

**Superconductivity in the H-layer of 4Hb-TaSSe**

Figure 3a shows a representative, high-resolution d*I*/d*V* curve acquired on the H-layer. The low-energy quasiparticle spectrum of this polymorph is dominated by a clear superconducting gap with large coherence peaks and in-gap features, i.e., a finite conductance with symmetric shoulders with respect to $E_F$ (red arrows). Our statistical analysis yields a superconducting gap $\Delta_H = 0.46 \pm 0.05$ mV (peak-to-peak, see supplementary note 4). Following the same procedure carried out for the T-layer, we also performed AR spectroscopy in the H-layer, which confirms the superconducting nature of the gap, as shown in Fig. 3b. Here, the evolution of the AR within the gap also falls beyond the simple isotropic BTK model (supplementary note 8), since a central AR peak emerges as the tip approaches the H-layer. The detection of AR processes in the conductance spectra is also confirmed by suppressing the Andreev states with magnetic fields (Fig. S5b).

We now analyze the in-gap spectroscopic features of the H-layer. First, the finite conductivity is tentatively attributed to the coupling between the two superconducting layers, as the energy positions of the shoulders at ±0.25 mV coincide with the coherence peaks of the superconducting gap of the T-polymorph in the layer underneath. This will be further corroborated in our calculations below. Second, the conductance at zero-energy always shows a finite value. However, it is not spatially constant throughout the H-layer as it maximizes at the edges (Fig. 3c). Figure 3d shows a series of d*I*/d*V* spectra acquired with a metallic (non-superconducting), calibrated tip across a T/H step edge (green line in Fig. 3c). The gap of the H-layer is uniform in bulk, slightly shrinking within ≲ 20 nm away from the edge. In this region, the zero-bias conductance is enhanced near the edge and gradually decreases with distance from the step-edge, eventually stabilizing at a constant value in the bulk (Fig. 3e). Moreover, the spectra evolve smoothly across the interface, showing a continuous connection between the in-gap shoulders in the H-layer and the superconducting gap in the T-layer (Fig. 3d). We estimate a decay rate of the zero-bias conductance of $d = 17 \pm 2$ nm (Fig. S9). This decay is not present in the normal state (Fig. S10), confirming that its origin is linked to superconductivity. A similar phenomenology was recently found in 4Hb-TaS$_2$, which was interpreted as a boundary mode in the framework of topological nodal-point superconductivity[13].

**Temperature and magnetic field dependence of superconductivity**

The initial electronic characterization of 4Hb-TaSSe uncovers superconductivity in both the T- and H-layers, each exhibiting significant differences in the size and shape of their superconducting gaps as well as in the behaviors of their corresponding Andreev reflections. These contrasting properties, occurring in the same material, raise intriguing questions about their effective coupling. To better understand this phenomenon, we measured the temperature and out-of-plane magnetic field below which each superconducting gap develops. Figures 4a,b show two representative sets of d$I$/d$V$ spectra displaying the evolution of the superconducting gap in the T and H layers with temperature ($T$) and $H_\perp$ field, respectively. As expected, the gap gradually reduces in both layers with increasing $T$ and $H_\perp$, and eventually closes at $T_C$ and $H_{C2}$. However, the critical parameters obtained from these and similar sets of measurements are significantly different in each layer.

Figures 4c,d summarize, respectively, the evolution of the superconducting gap size extracted from the data sets in Figs. 4a,b as a function of $T$ and $H_\perp$ for the T- (red dots) and H- (blue dots) layers. The colored, horizontal lines on the x-axis represent the statistical uncertainty of the critical values extracted from the different datasets. As can be observed, the gaps gradually decrease with increasing $T$ in both layers and appear to vanish for each layer at $T_{C,T} = 1.3 \pm 0.2$ K and $T_{C,H} = 2.2 \pm 0.2$ K, following the BCS behavior of uncoupled gaps rather well. The larger $T_{C,H}$, coincides with the critical temperature extracted from our magnetization, transport and specific heat measurements on 4Hb-TaSSe (supplementary note 11). We emphasize that none of our temperature-dependent d$I$/d$V$ datasets showed a "*tail*" in $\Delta_T$(T) for $T_{C,T}<T<T_{C,H}$. Therefore, $T_{C,H}$ surpasses in nearly 1 K $T_{C,T}$ which after taking into account the standard deviation of the measurements (shaded areas along the $T$ axis in Fig. 4c), suggesting weakly coupled condensates in each layer that seem effectively decoupled above $T_{C,T}$.

The $H_\perp$-field dependence is even more intriguing. As seen in Fig. 4d, the gaps on each layer show very different evolution with the field at T = 0.34 K, vanishing in each case at $H_{C2,T} = 2.6 \pm 0.3$ T and $H_{C2,H} = 1.0 \pm 0.1$ T. Remarkably, the smaller gap in the T layer survives up to 2.6 times larger fields than the larger gap in the H layer. This behavior was found regardless of the polarity (field direction) of $H_\perp$ on both T- and H-polymorphs (supplementary note 12). The large difference between the two critical fields (~1.6 T), much larger than the experimental uncertainty and standard deviation, is consistent again with a scenario of effective decoupling of the gaps in each polymorph above $H_{C2,H}$. We note that we used different fitting methods (s-wave and nodal) to obtain the Δ values plotted in Fig. 4 but no significant effect was found and the conclusions remain valid. Lastly, for completeness, we tracked the evolution of the superconducting gaps in both layers as a function of temperature for finite $H_\perp$ fields. The extracted $T_C = T_C (H_\perp)$ values measured in each layer are shown plotted in Fig. 5.

**DFT band structure and effective multigap model**

To understand the potential emergence of a layer-resolved multigap superconducting state, we perform DFT calculations of the band structure of 4Hb-TaSSe in the $\sqrt{13}\times\sqrt{13}$ CDW state, using a crystal structure based on 4Hb-TaS$_2$ where 50% of randomly chosen S atoms are replaced by Se. In an isolated T layer, the CDW distortion gives rise to a half-filled flat band at the Fermi level which leads to a Mott insulator when correlations are included[30–32]. In bulk 4Hb compounds, however, the presences of interlayer hopping and charge transfer lead to a significant $k_z$-dispersion and carrier depletion of the T-derived band[33–35]. This is also the case in 4Hb-TaSSe, as seen by color coding the bands according to their weights in the T- and H-layers, as shown in Fig. 6a. To better quantify these features, we compared the DFT bands to a tight-binding model comprising the metallic band of the H-layer folded back onto the smaller BZ of the T-layer and coupled to a single T-layer orbital representing the flat band localized at the SoD center[36] [supplementary note 8]. As seen in Fig. 6b, this model reproduces the decoupled narrow band seen around 250 meV at $k_z = 0$, as well as its dispersion and mixing with the H bands as the Γ-A line is traversed. The obtained interlayer tunneling by this fit is $t_\perp = 90$ meV, the same order of magnitude as the one obtained in H-T bilayers[36,37]. The layer-projected DFT Fermi Surfaces are shown in Figs. 6c,d, where it is observed that despite a significant charge transfer, there is sizable weight of the flat band state in some portions of the Fermi Surface. This is in line with previous DFT calculations which show that the flat band is less depleted in 4Hb-TaSe$_2$ compared to 4Hb-TaS$_2$ (ref. [33]).

With STM and DFT suggesting a sizable interlayer hybridization (Fig S2, Fig 6b) and the presence of quasiparticles from both T- and H-layers at the Fermi level (Figs. 1e,f and Figs. 6c,d), we develop a minimal bilayer model that captures the essential low-energy degrees of freedom to solve the coupled superconducting gap equations (see Methods). The model consists of light and heavy bands with independent in-plane dispersions for the H and T states, respectively, and a $k_z$-dependent hybridization that is maximum at $k_z = 0$ and vanishes at the BZ boundaries $k_z = \pm 2\pi$, as imposed by the symmetries of the bilayer. This hybridization between the T- and H-layers results in the quasi-2D multi-band Fermi surface shown in Fig. 6e, with weakly dispersing cylinders centered at Γ and K. As in DFT, the K centered pocket shows a dominant T-character at $k_z = 0$ which weakens towards the BZ boundary ($k_z = \pm 2\pi$); the Γ pocket has a prevalent H-character instead. The model also contains independent intra-layer uniform attractive pairing interactions with $V_H > V_T > 0$; although our STM experiments suggests that one of the gaps is nodal, for simplicity we assume s-wave gaps in each layer. The $k_z$-dependent layer-character of the FS pockets results in anisotropic gap structures with low values of the gap in areas of the FS with strong T-character (Fig. 6f). Moreover, the resulting DOS projected on the H-layer develops in-gap features corresponding to the coherence peaks of the T layer (blue curve in Fig. 6g), very similar to those measured in d$I$/d$V$ (Fig. 3a), and often seen in multigap superconductors[38]. Despite the absence of inter-layer pairing in the model, the $k_z$-dependent interlayer hybridization also opens up a gap in the T-layer projected DOS, even in the absence of attraction in that layer ($V_H > 0, V_T = 0$, see Fig. S14). As a result, the peak-to-peak distances of the T-layer local DOS are generally larger than $2\Delta_T$ at low temperatures (red curve in Fig. 6g), surviving up to the temperature at which $\Delta_H \to 0$. This induced gap in the T-layer is not seen experimentally in our d$I$/d$V$ curves for $T_{C,T} < T < T_{C,H}$ (Fig. 4c), suggesting that additional effects not included in the model lead to an effective decoupling between the gaps in this temperature regime.

**Discussion**

The local electronic structure of the T- and H-layers exhibit a finite DOS, which enables the development of superconductivity in both layers, as revealed by our STS measurements. This metallic character, however, differs in each layer. While the H-layer DOS is homogeneous, the T-layer DOS shows marked spatial variations due to the presence of a zero-bias resonance at the center of roughly half of the SoDs (Fig. 1f-h). This resonance, previously observed in T/H heterostructures of Ta- and Nb-based TMDs[36,39–43], has been interpreted either as a Kondo peak, which would arise from the screening of SoD magnetic moments by the conduction electrons from the metallic H-layer, or as a quasiparticle peak from a doped Mott insulator[37], where the role of the H-layer is to provide electron charge transfer. In bulk 4Hb compounds, where this peak has also been observed[40], neither of these scenarios is likely to apply since according to our calculations (Fig. 6a and supplementary note 13) the flat band acquires a bandwidth of nearly 300 meV, clearly larger than the Hubbard band splitting observed in monolayers. While further analysis is required to establish the origin of the zero bias peak, the last surface layer might have different interlayer hopping and occupation compared to the bulk due to surface potential effects, bringing it closer to a more correlated regime. This may also make it more susceptible to chalcogen disorder and explain the peak variability. Nevertheless, we emphasize that the superconducting gap in the T-layer appears to be independent of these zero-bias peak fluctuations, which occur in much shorter length scales than the coherence length.

The observed effective decoupling of the T- and H-layer superconducting gaps is particularly remarkable considering the substantial T-H interlayer tunneling, and is further supported by the distinctly different magnetic field responses of each gap (Fig. 5), which supports the framework of layer-selective superconductivity. In most cases, a magnetic field tends to suppress the smaller gap first, and the upper critical field is then typically determined by the large gap[44]. However, if the T-layer is deeper into the dirty limit than the H-layer, its small gap could determine $H_{C2}$ at low temperatures[2]. Moreover, the in-plane Fermi velocity in this system is much larger (smaller) in areas with strong H-character (T-character) electrons (Fig. 6h), which could result in stronger orbital effects for electrons in the H-layer, and hence a faster suppression of superconductivity. In fact, while the gap in the H-layer closes at fields much lower than its BCS paramagnetic limit, $H_{P,H}$ = *4.1* T, suggesting strong orbital effects, the gap in the T-layer seems to be Pauli limited instead, $H_{P,T}$ = 2.4 T (Fig. 5).

Our work reveals a fundamentally novel form of multiband superconductivity in 4Hb-TaSSe, characterized by effectively spatially decoupled superconducting condensates— a phenomenon not observed in previously known systems. In conventional multigap superconductors like MgB$_2$ or Fe-based compounds, multiple superconducting gaps arise from different Fermi surfaces but coexist in real space, preventing their independent external control. In contrast, 4Hb-TaSSe unlocks this latent degree of freedom: the superconducting condensates reside in distinct polymorphic layers (T and H), each exhibiting different gap structures and critical responses to temperature and magnetic field. This spatial separation and weak interlayer coupling enable independent tuning of the condensates by magnetic fields and temperature, establishing a new paradigm for layered superconductivity. Our findings in 4Hb-TaSSe open new avenues for engineering layer-selective, tunable superconducting devices using transition metal dichalcogenide polymorphs—offering unprecedented control over individual superconducting states at the nanoscale.

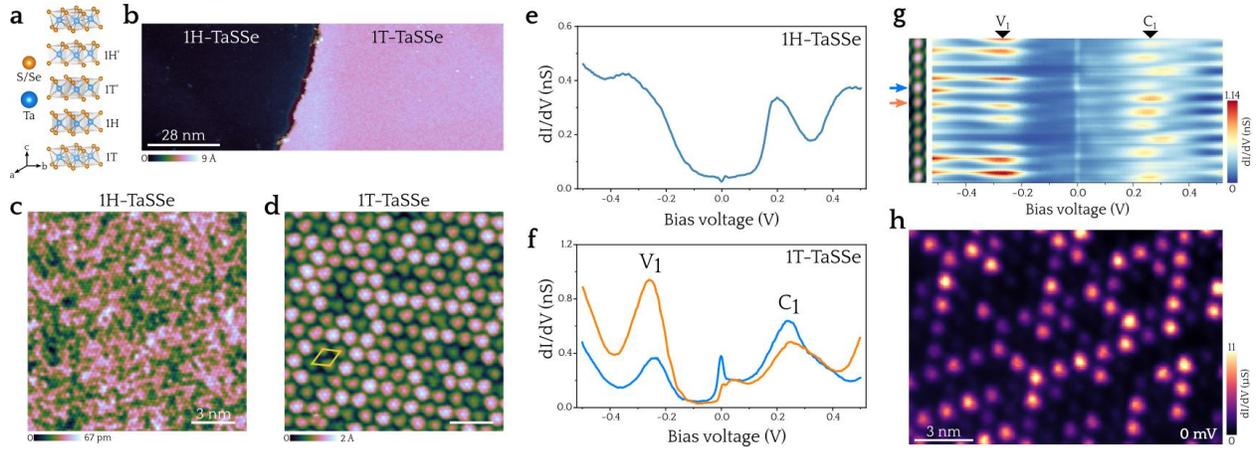

**Fig. 1 - Electronic structure of the 4Hb-TaSSe polytype. a,** Sketch of the 4Hb-TaSSe unit cell. **b**, Large-scale STM image showing the two consecutive T- and H-polymorph layers ($V_s$ = -0.5 V, $I$ = 40 pA, $T$ = 0.34 K). **c,d**, Atomically resolved STM images of the H-layer ($V_s$ = 0.2 V, $I$ = 0.2 nA, $T$ = 4.2 K) and T-layer ($V_s$ = 90 mV, $I$ = 20 pA, $T$ = 0.34 K), respectively. In the latter, the atomic registry is overlaid by the $\sqrt{13}\times\sqrt{13}$ CDW (unit cell depicted in yellow). **e**, d$I$/d$V$ spectrum acquired in the H-layer ($T$ = 4.2 K). **f**, d$I$/d$V$ spectra measured on the center of two differentiated SoDs indicated by arrows in **g**. **g**, d$I$/d$V$ ($x,V$) map recorded along the STM topography depicted on the left, which follows a high-symmetry direction of the CDW ($T$ = 4.2 K). **h**, Differential conductance (d$I$/d$V$) map acquired at $E_F$ ($V_s$ = 0 V) ($T$ = 4.2 K).

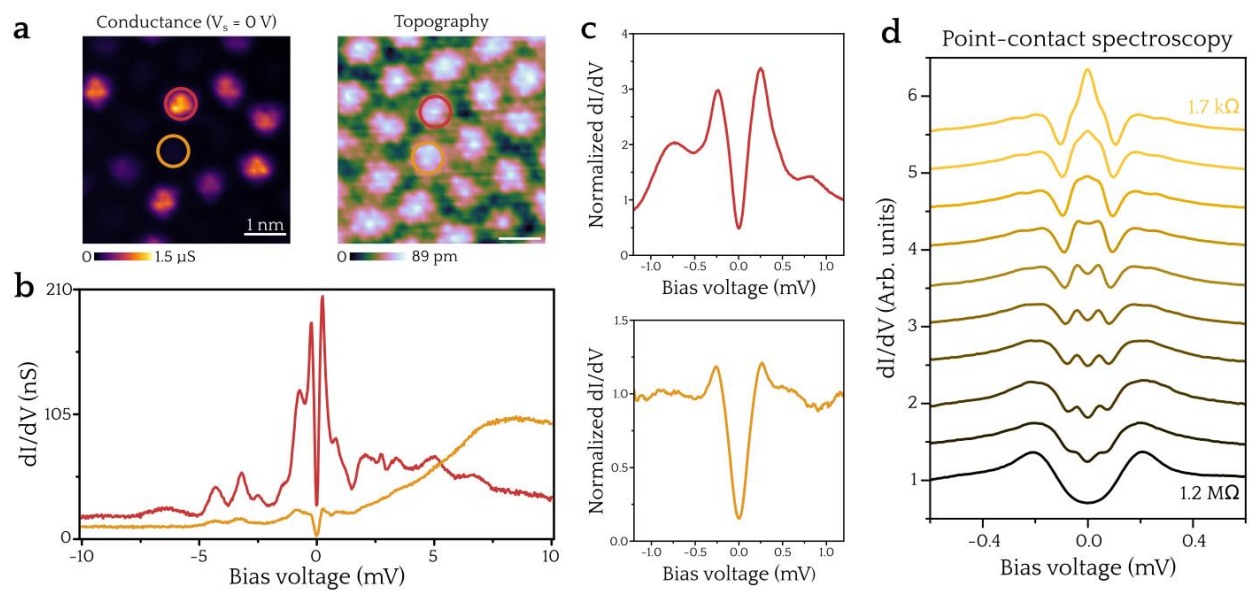

**Fig. 2 - Superconductivity in the T-polymorph. a**, Zero-energy d$I$/d$V$ map and corresponding STM topography of the same area ($V_s$ = 0.6 V, $I$ = 50 pA). **b**, Representative d$I$/d$V$ spectra recorded at the center of the SoD clusters encircled in **a** using the same colors. **c**, Normalized high-resolution d$I$/d$V$ spectra measured in the two contrasted CDW-SoD revealing the superconducting gap. **d**, Set of Andreev reflection spectroscopy spectra acquired consecutively with a metallic (non-SC) tip upon reducing the tunneling junction resistance. All data were measured at $T$ = 0.34 K.

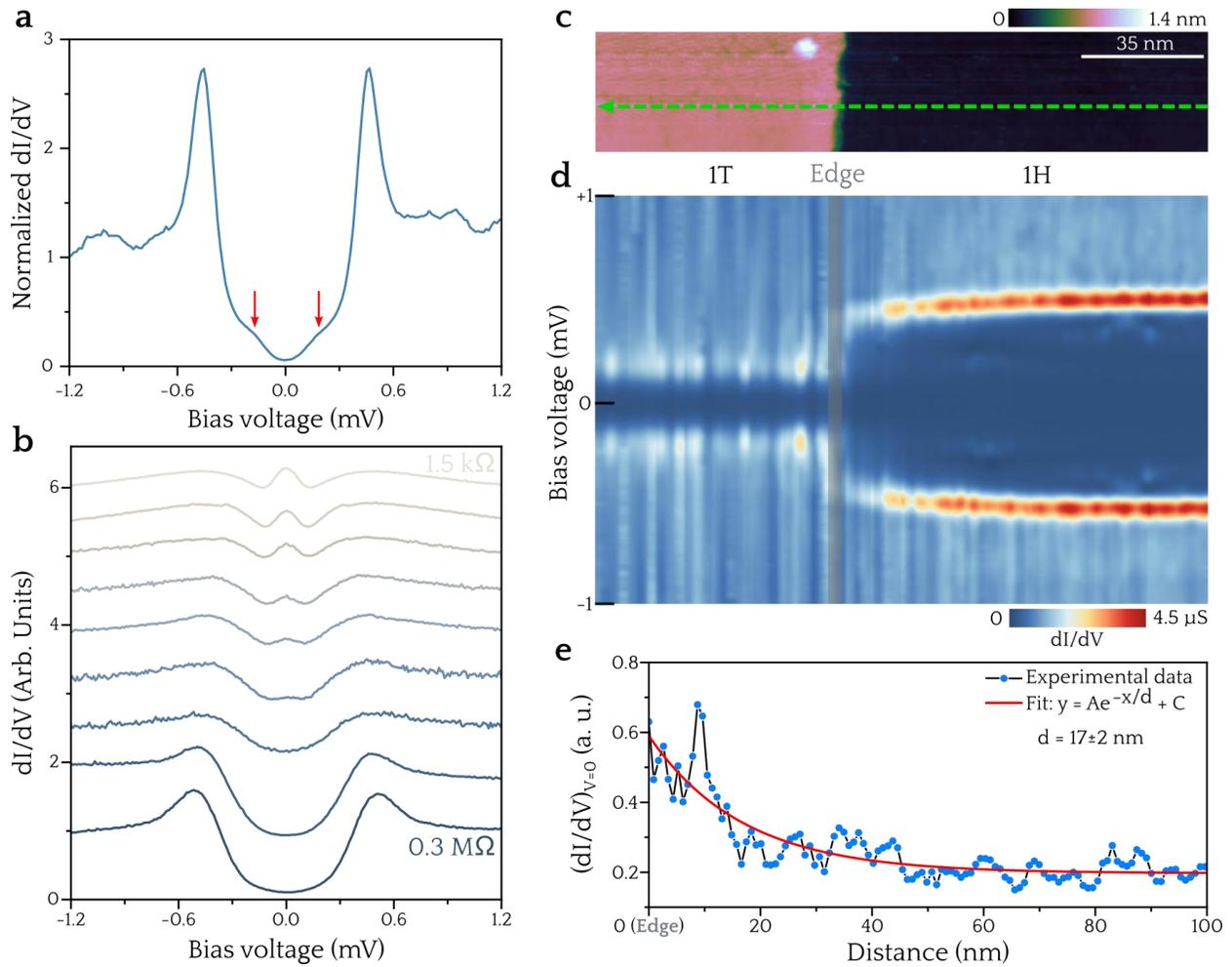

**Fig. 3** - **Superconductivity in the H-polymorph. a**, Normalized d$I$/d$V$ spectrum showing the superconducting gap structure of the H-layer. Red arrows indicate in-gap finite conductance shoulders. **b**, AR spectroscopy spectra acquired at different tunneling junction resistance. **c**, STM topograph showing two consecutive T- and H-type layers ($V_s$ = 0.1 V, $I$ = 20 pA). **d**, High-resolution tunneling d$I$/d$V(x,V)$ taken along the dashed green line in **c**. The gray rectangle indicates the location of the atomic step edge. **e**, Zero-energy differential conductance profile, d$I$/d$V(x, V_s = 0)$, extracted from **d**. The red line shows the fitting results of a first order exponential decay. All data were measured at $T$ = 0.34 K.

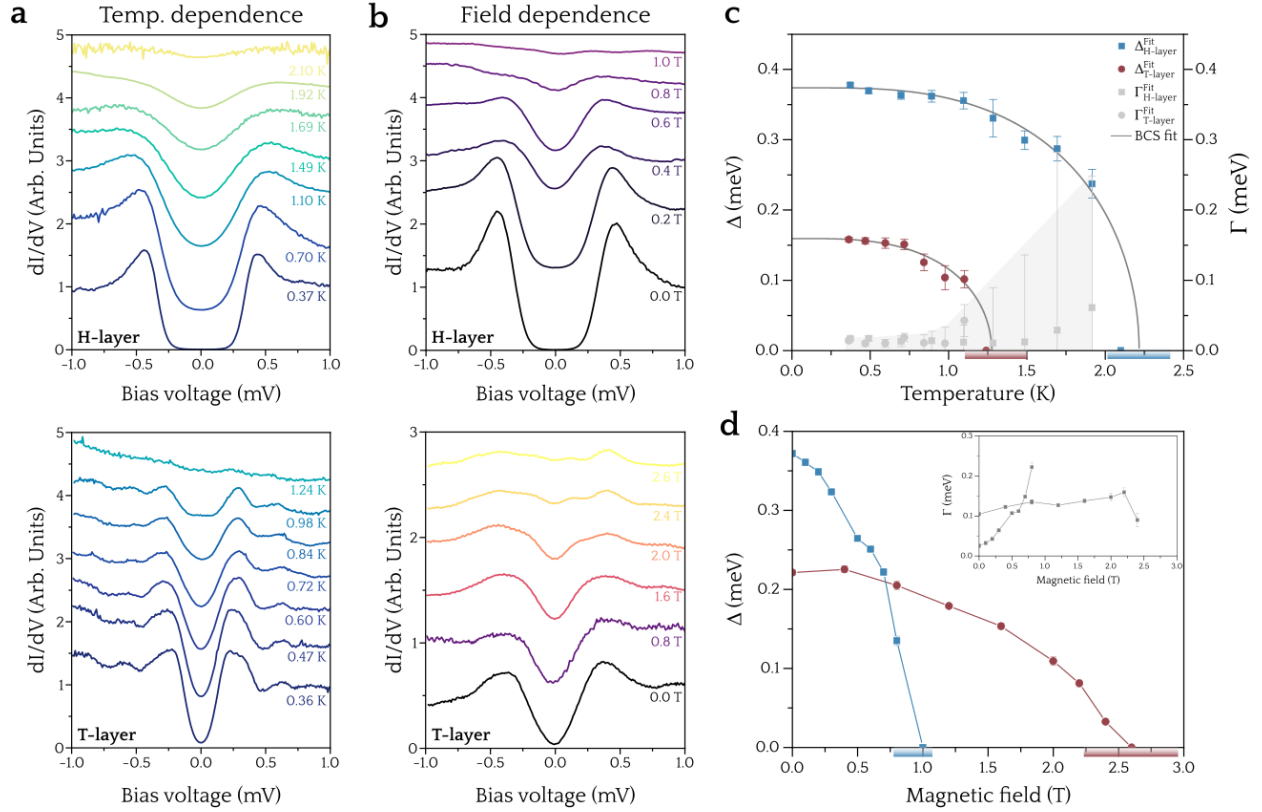

**Fig. 4** - **Temperature and magnetic field dependence of the superconducting state of the polymorph layers.** Representative dI/dV spectra sets showing the temperature (**a**) and magnetic-field strength (**b**) evolution of the superconducting gap on the H (upper panel) and T (lower panel) layers. STS data in **b** were measured at $T = 0.34$ K. Superconducting gap, $\Delta$, and quasiparticle lifetime broadening, $\Gamma$, as a function of **c**, temperature and **d**, magnetic field ($\Gamma$ in the inset). The $\Delta$ and $\Gamma$ values were extracted by fitting the d$I$/d$V$ spectra to the Dynes function (see supplementary note S6). The error in $\Gamma$ becomes $\sim\Delta$ near $T_C$ due to increased fit uncertainty. The elongated colored shaded areas mark the statistical uncertainty of $T_C$ and $H_{C2}$ extracted from all the measured sets.

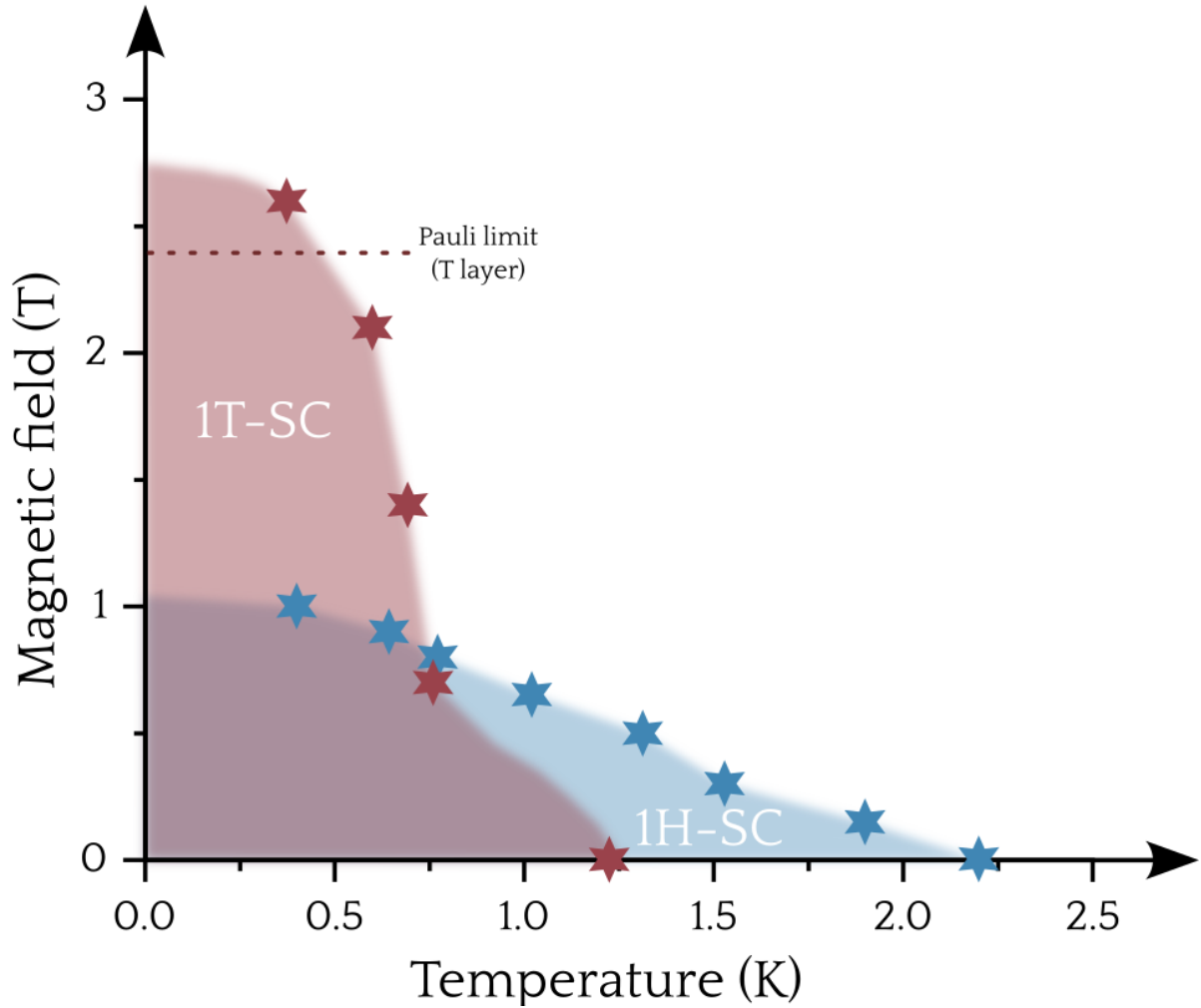

**Fig. 5** - **Superconducting regimes in the distinct polymorph layers.** Schematic representation of the superconducting regimes in each polymorph in 4Hb-TaSSe. Blue and red stars represent the critical values experimentally obtained for the T- and H-layer, respectively. The shaded areas are drawn for illustration purposes.

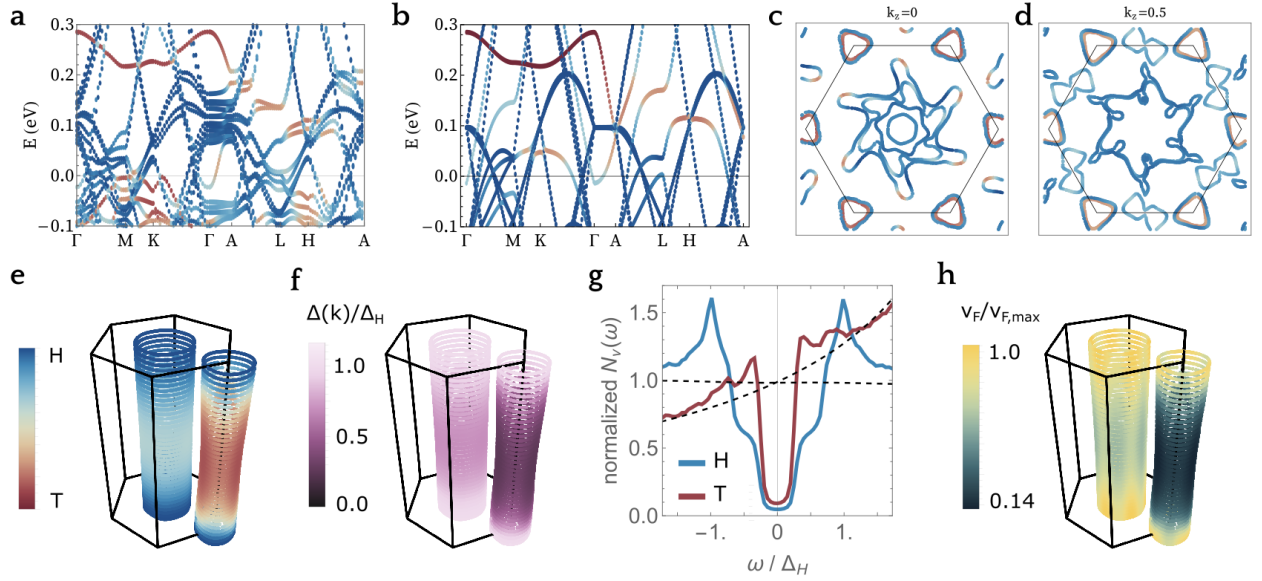

**Fig. 6 - DFT band structure and low energy model. a**, DFT band structure of 4Hb-TaSSe in the √13×√13 CDW state, color coded according to weight of the wavefunction in the H (blue) or T (red) layer. **b**, Band structure of the tight-binding model described in the text. **c**, Fermi surface corresponding to the DFT band structure for momentum $k_z = 0$. **d**, Same for $k_z = \frac{\pi}{c}$. **e**, Fermi surface of the low-energy model, color coded with the weight of the coupled H and T layers. **f**, Superconducting gap for onsite intralayer attraction ($\Delta_T = 0.11\Delta_H$). **g**, Corresponding local DOS projected on the H layer (blue) and the T layer (red). The dashed lines denote the normal state local DOS. **h**, In-plane Fermi velocity, which correlates with the T/H character of the bands in **e**.

## Methods

### 4Hb-TaSSe crystal growth

Single crystals of 4Hb-TaSSe were grown using a standard chemical vapor transport method. A stoichiometric amount of high purity powder of Ta, Se and S were sealed together in a quartz tube along with Iodine as a transport agent. Slightly excess amounts of sulfur and selenium were taken to ensure the desired composition. After that, the sealed tube was positioned inside a two-zone furnace with growth zone temperature 820 °C for a duration of 12 days. Millimeter-sized single crystals were successfully obtained at the cooler end of the tube.

### STM/STS experiments

Experiments were performed on a UHV system that allocates a commercial STM (USM1300, Unisoku) with capabilities to operate at 0.34 K and under high magnetic field (11 T, orthogonal to the sample surface). All the experimental data were recorded in the temperature range of 0.34 K - 4.2 K using Pt/Ir tips, which were pretreated on Au(111) or Cu(111) and calibrated against their respective Shockley surface states. Standard lock-in technique was employed for STS data acquisition, in which a peak-to-peak AC modulation voltage ($V_{a.c.}$, detailed in each figure) at $f$ = 833 Hz was applied to the sample bias voltage during the acquisition. Typical *a.c.* modulation voltages of 6-30 µV were employed for high-resolution STS and 1 mV for large-scale STS. All STM/STS data were post-processed and analyzed employing the freeware WSxM[45]. In order to ensure high quality surfaces with low defect concentration, the 4Hb-TaSSe crystals were mechanically exfoliated under UHV conditions in the load-lock chamber ($2\times10^{-9}$ mbar) using tape that was previously attached on top of the surface in air. After exfoliation, the samples were immediately transferred into the STM for measurements.

### DFT calculations and effective model

All DFT calculations were carried out using Vienna *ab-initio* Simulation Package (VASP)[46,47] v.6.2.1. with projector-augmented wave pseudopotentials using the General Gradient Approximation with Perdew Burke Ernzerhof parametrization (PBE)[48]. Self-consistent calculations were found to be well converged with a 480 eV kinetic cutoff and a gamma-centered 13×13×3 k-mesh. The 4Hb-TaSSe structure used in DFT is built from 4Hb-TaSe$_2$ in its $\sqrt{13} \times \sqrt{13}$ CDW[49] by replacing 50% of Se atoms by S atoms in randomly chosen positions[36]. Projected band structures from Fig. S4 were obtained using the PyProcar package in Python[50].

To extract interlayer hybridization, we set up a tight binding model generalizing the one in Ref. [36] to a bulk structure. This model contains a single band for the H layer, fitted to a monolayer calculation and then folded onto the CDW unit cell induced by the T layer (which results in 13 bands). An extra orbital for the T layer located at the SoD center represents the flat band with intralayer hopping $t_{SoD}$ and on-site energy $E_{SoD}$. This state is hybridized with the H layer states via an interlayer hopping $t_z$. Since it is known that states in the H layer near $\Gamma$ are made mostly of out-of-plane $d_{z^2}$ orbitals, while states at the BZ edge are mostly made of in-plane orbitals $d_{x^2-y^2}$ and $d_{xy}$, tunneling is expected to be stronger to states near $\Gamma$, which we encode in an effective way in a momentum-dependent tunneling $t_\perp e^{-A|\vec{k}|/|\vec{G}|}$ where $\vec{G}$ is a reciprocal lattice vector of the CDW BZ. The parameters in the model are chosen so that the overall bands match the ones obtained by DFT, and take values $t_{SoD}$= 15 meV, $t_z$= 90 meV, $t_{SoD}$= 240 meV, A = 2.

This model reproduces the dispersion and hybridization of the flat band, particularly in the $k_z$ direction. Note that this model has an effective translation symmetry because the H, H' and T, T' layers are represented as equivalent, so the actual unit cell is only a single T/H bilayer. The hybridization between H and T bands is strongest at $\Gamma$ and actually vanishes at the edge of the enlarged BZ, $k_z = 2\pi/c$. When the bands are folded back onto the physical unit cell, this produces two copies of the flat band (essentially bonding and antibonding combinations), with the one folded from $k_z = 2\pi/c$ remaining completely unhybridized. This effect is clearly seen in the DFT calculation, providing support for our tight binding parametrization.

To address the pairing problem, we take input from the previously obtained band structure to develop a minimal bilayer model that captures the essential degrees of freedom near the Fermi level, containing a single T/H bilayer and a single band per layer

$$\mathcal{H}_0 = \sum_k \left( \xi_H(k_x, k_y) c_{Hk}^\dagger c_{Hk} + \xi_T(k_x, k_y) c_{Tk}^\dagger c_{Tk} + 2t_z \cos\frac{k_z}{4} c_{Hk}^\dagger c_{Tk} \right) + h.c.$$

with intralayer energy dispersions $\xi_{H/T}(k_x, k_y) = t_{1,H/T}[\cos k_x + 2\cos(k_x/2)\cos(\sqrt{3}k_y/2)] + t_{2,H}[\cos(\sqrt{3}k_y) + 2\cos(3k_x/2)\cos(\sqrt{3}k_y/2)] - \mu$ in a trigonal lattice, and an interlayer hopping $t_z$ connecting the H and T layers along $k_z$. The hopping amplitudes and chemical potential µ are taken to match the overall behavior of the band structure ($t_{1,H} = 87$, $t_{2,H} = 215$, $t_{1,T} = 15$, $t_z = 20$ and $\mu = -34$, all in units of meV) and result in a three-dimensional Fermi surface consisting of weakly dispersing cylinders centered at the $\Gamma$ and K points in the BZ, with mixed H and T electronic character (Fig. 6e). We consider on-site intralayer attraction in the s-wave Cooper channel, and self-consistently solve the coupled gap equations with solutions of $\Delta_\nu = V_\nu \sum_k <c_{\nu-k\downarrow} c_{\nu k\uparrow}>$ for $\nu$ = H, T layer index and $V_H > V_T > 0$.

**Acknowledgements**

We thank Kevin Nuckolls and Mathias Scheurer for useful discussions. M.M.U. acknowledges support by the ERC Starting grant LINKSPM (Grant #758558). M.M.U. and M.N.G. acknowledge support by the grant ID2023-153277NB-I00 funded by the Spanish Ministry of Science and Innovation. M.N.G is also supported by the Ramon y Cajal Grant RYC2021-031639-I funded by MCIN/AEI/ 10.13039/501100011033 and EU NextGenerationEU/PRTR. H.G. acknowledges funding from the EU NextGenerationEU/PRTR-C17.I1, as well as by the IKUR Strategy under the collaboration agreement between Ikerbasque Foundation and DIPC on behalf of the Department of Education of the Basque Government. F.J. is supported by grant PID2021-128760NB-I00 funded by the Spanish Ministry of Science and Innovation and by the Leonardo fellowship from BBVA. R.M.F. was supported by the Air Force Office of Scientific Research under Award No. FA9550-21-1-0423. M.G.V received financial support from the Canada Excellence Research Chairs Program for Topological Quantum Matter, Diputación Foral de Gipuzkoa Programa Mujeres y Ciencia and PID2022-142008NB-I00 projects funded by MICIU/AEI/10.13039/501100011033 and FEDER, UE. T.A. acknowledges the Department of Science and Technology (DST), Government of India, for financial support through Award No. DST/INSPIRE/03/2021/002666). R. P. S. acknowledges the SERB, Government of India, for the Core Research Grant No. CRG/2023/000817. S.S. and I.S.R. acknowledge enrollment in the doctorate program "Physics of Nanostructures and Advanced Materials" from the "Advanced polymers and materials, physics, chemistry and technology" department of the Universidad del País Vasco (UPV/EHU).

**Author contributions**

M.M.U. conceived the project. H.G. and S.S. measured the STM/STS data with the help and M.M.U.. H.G. analyzed the STM/STS under the supervision of M.M.U.. T.A. and C.P. carried out the sample growth and bulk crystal characterization under the supervision of R.P.S. I. R.S. performed the ab-initio calculations under supervision of M.V.. F.J. developed the bulk TB model and fits to *ab initio* calculations. M.N.G. and R.M.F. developed the effective model for superconductivity. A.B.P solved the gap equations and spectroscopic features of the effective model under the supervision of M.N.G.. H.G., M.M.U. and M.N.G. wrote the manuscript. All authors contributed to the scientific discussion and manuscript revisions.

Supplementary Materials for

# Layer-selective Cooper pairing in an alternately stacked transition metal dichalcogenide

Haojie Guo, Sandra Sajan, Irián Sánchez-Ramírez, Tarushi Agarwal, Alejandro Blanco Peces, Chandan Patra, Maia G. Vergniory, Rafael M. Fernandes, Ravi Prakash Singh, Fernando de Juan, Maria N. Gastiasoro, Miguel M. Ugeda

**Supplementary Note 1: Powder and single-crystal XRD characterization of 4Hb-TaSSe**

The identification of the polytype 4Hb-TaSSe was conducted using powder X-ray diffraction (XRD) at room temperature. The refined diffraction pattern of crushed crystal powder confirmed a hexagonal structure categorized in the P6$_3$/mmc (194) space group. The obtained lattice parameters are $a = b = 3.393 \pm 0.002$ Å, and $c = 24.608 \pm 0.002$ Å. The presence of high intensity and sharp peaks in the single crystal diffraction pattern validates the exceptional crystallinity of the sample, a conclusion further reinforced by the Laue pattern (inset of Fig. S1a). The recorded peaks along the [00l] planes in the single crystal XRD pattern indicate that the $c$-axis is perpendicular to the crystal plane (Fig. S1b). Furthermore, from energy-dispersive X-ray spectroscopy (EDS) measurements, we determined that the atomic ratio of Ta:S:Se closely approximates 1:0.99:0.9.

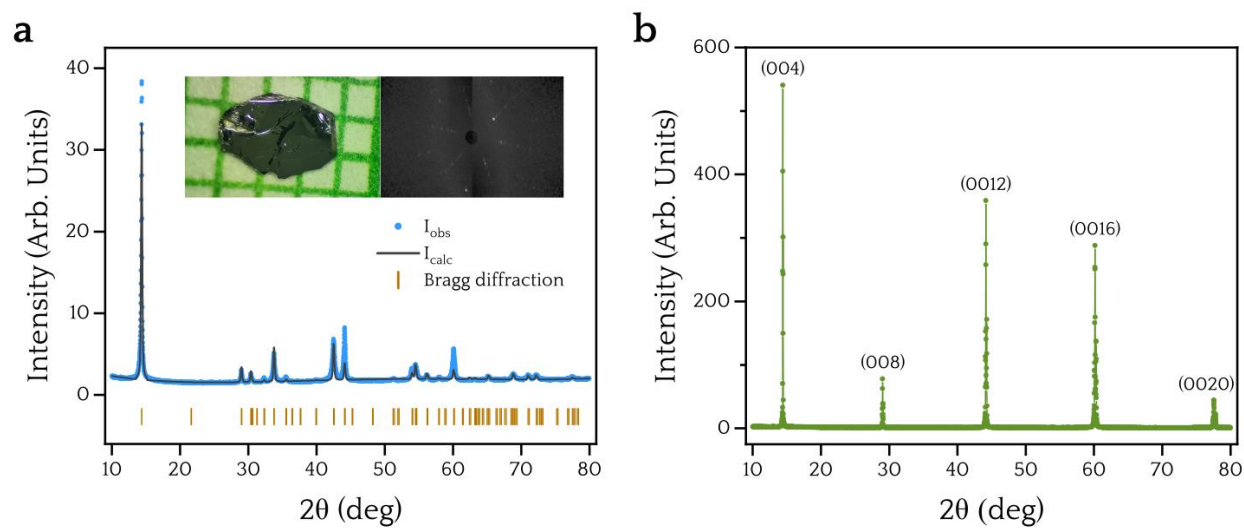

**Fig. S1 - Crystal structure characterization of 4Hb-TaSSe. a**, Crystal powder XRD pattern. Insets show the Laue pattern and a photograph of the crystal. **b**, Single-crystal XRD pattern with peaks along the (00l) direction.

**Supplementary Note 2: Absence and transparency of the CDW of the H-polymorph layer**

In Fig.1c of the main text, we present an atomically resolved STM topography of the H-layer, which clearly shows the absence of a CDW on its surface. However, in this type of STM images, as seen in Fig. S2a, we routinely observe the underlying √13×√13 CDW of the T-layer through the H-layer. The inset displays the 2D-FFT of the topography image, highlighting both the 1×1 spots of the H-layer atomic lattice and the spots corresponding to the T-layer CDW underneath. Spatial localization of these structures is achieved by applying high-pass filtering to isolate the respective Bragg spots from the FFT (Fig. S2b).

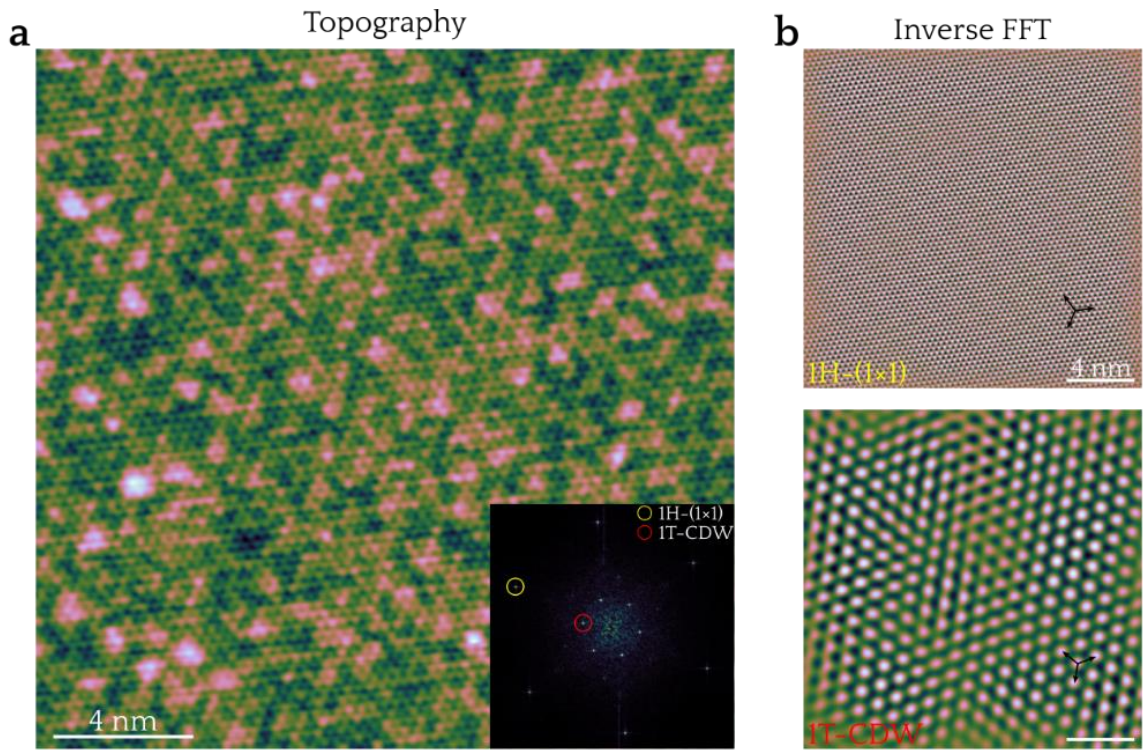

**Fig. S2 - Real-space visualization and FFT decomposition of the H-polymorph. a**, Atomically resolved STM topography image of H-layer. The inset shows the 2D-FFT of the topography image ($V_s$ = 30 mV, $I$ = 0.2 nA, $T$ = 0.34 K). **b**, Inverse FFT images obtained by subtracting the signals from the Bragg spots related to the H-layer (1×1) and the T-layer CDW, respectively. Arrows indicated the high-symmetry directions of the hexagonal lattice.

**Supplementary Note 3: Spatially dependent electronic structure of T-layer**

The large-window electronic structure of the T-polymorph layer shows a spatial dependence across the surfaces, as shown in the main text (Fig. 1f-h). In particular, the intensity of the main peaks in the STS spectra exhibits a varying intensity modulation when measured across different SoD clusters, as observed in d$I$/d$V(x,V)$ maps (Fig. 1g). We further corroborated this by acquiring d$I$/d$V(V)$ maps at relevant energies close to the main peak positions, as depicted in Fig. S3. We observe that the relative d$I$/d$V$ intensity of the SoD clusters varies with the acquisition energy. In particular, we observe anticorrelation in the d$I$/d$V$ intensity for occupied and unoccupied states, for instance, in the 6 SoD clusters highlighted in Fig. S3. Interestingly, the d$I$/d$V$ intensity at zero bias is directly correlated with the occupied states energies.

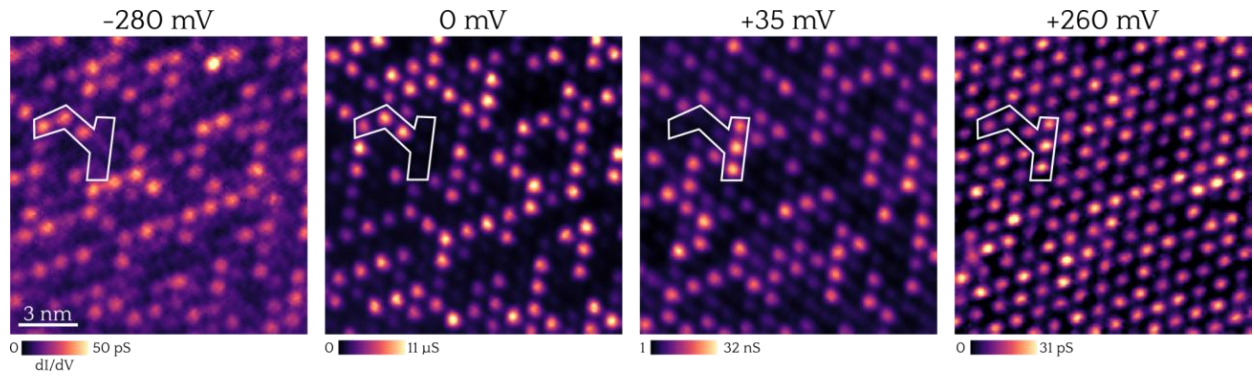

**Figure S3 - Spatial dependence of the electronic structure of the T-layer.** Differential conductance maps measured at the relevant energies in the STS spectra of T-termination. Six SoD clusters are marked in all images, highlighting the variation of the d$I$/d$V$ intensity with the bias voltage. Acquisition parameters: $V_{a.c.} = 3$ mV, $T = 4.2$ K.

**Supplementary Note 4: Statistical analysis of the superconducting gap size of the polymorphs**

Figure S4 shows the resulting histograms of the statistical analysis performed to estimate the superconducting gap values for each polymorph layer. Here, for simplicity, we define the gap as the energy distance between the coherence peaks in our d$I$/d$V$ spectra. For this statistical analysis, we chose statistically significant d$I$/d$V$ spectra obtained on different samples, regions, and/or with different tips. The results reveal an average superconducting gap size of $\Delta_T = 0.23 \pm 0.04$ meV and $\Delta_H = 0.46 \pm 0.05$ meV (standard deviation), for the T- and H-layers, respectively. Note that the plotted gap values in Figs. 4c,d are smaller because they are extracted by fitting the d$I$/d$V$ spectra to the Dynes function. Nonetheless, the ratio between them remains approximately constant.

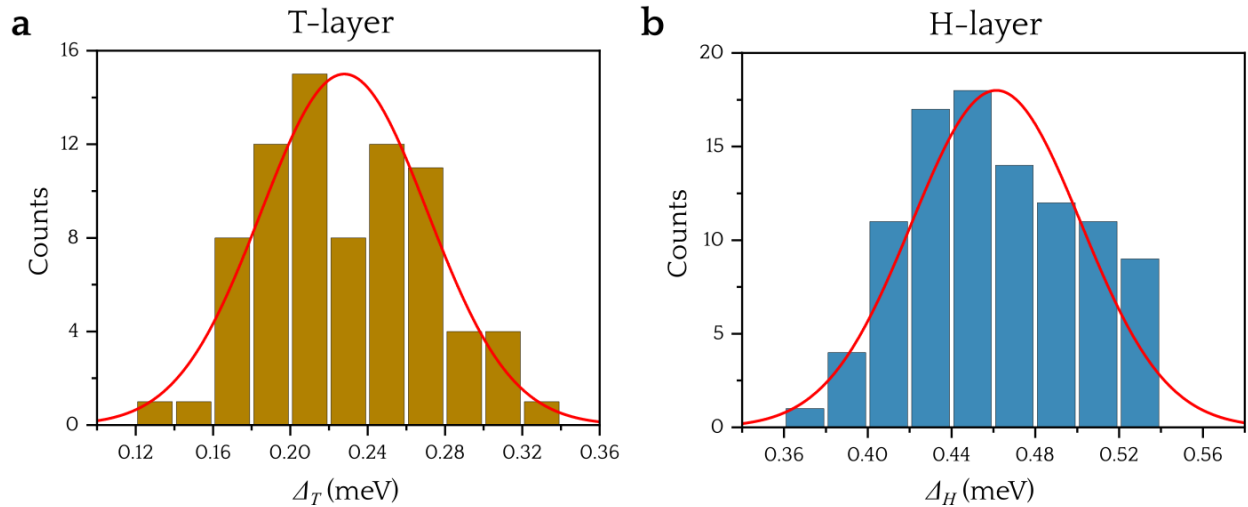

**Fig. S4 - Statistical analysis of the superconducting gap size in the polymorphs.** Histograms showing the energy distance between the coherence peaks in the T-layer (**a**) and H-layer (**b**). Only statistically significant d$I$/d$V$ spectra are considered. The red lines indicate the normal distribution.

**Supplementary Note 5: Magnetic field dependence of Andreev reflection spectroscopy measurements on T- and H-layers**

The observation of in-gap features in tip-sample height-dependent measurements in the T-layer (Fig. 2d) and the H-layer (Fig. 3b) is ascribed to the formation of Andreev states. Therefore, by lifting the superconducting state, such states should also disappear. To this end, we studied the evolution of these in-gap states for tunneling junction resistances of 1.7 k$\Omega$ (T-layer) and 0.6 k$\Omega$ (H-layer) with an out-of-plane magnetic field. The data are depicted in Fig. S5, where we observe in both cases the suppression of the Andreev states as $H_\perp$ is increased. The present results also set an upper limit for $H_{C2} = 3$ T on the T-layer, and $H_{C2} = 1.2$ T on the H-layer, which is in agreement with results derived from quasiparticle tunneling spectra data (Fig. 4c).

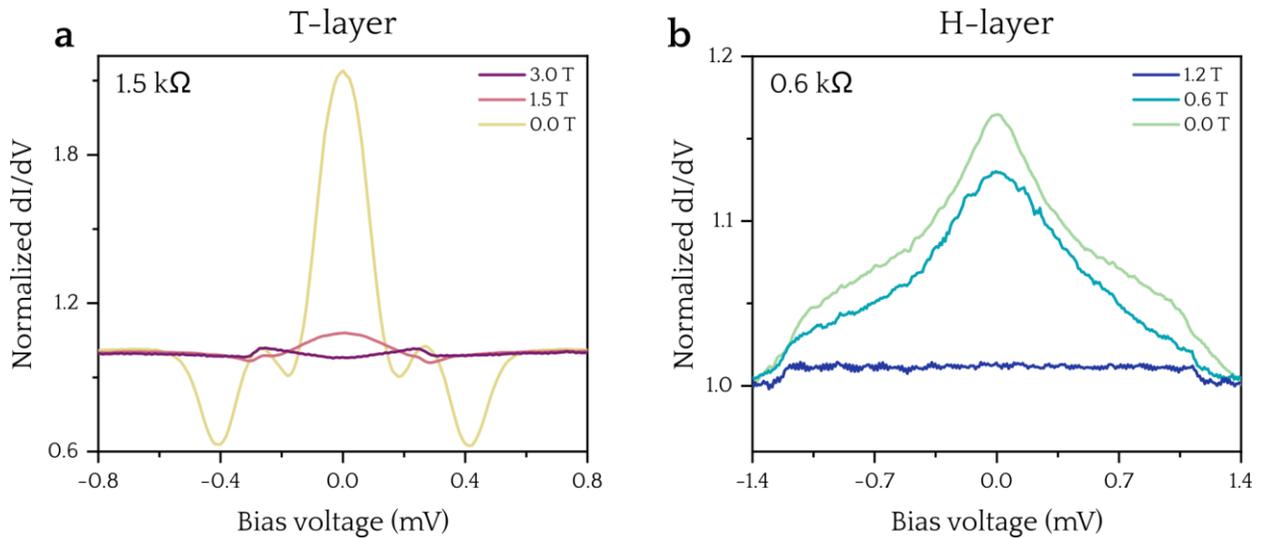

**Fig. S5 - Magnetic-field dependent Andreev reflection spectroscopy measurements. a,b**, Andreev reflection spectrum evolution as function of $H_\perp$ measured on the T- and H-polymorph layers, respectively. All the data shown here were measured at $T = 0.40$ K.

**Supplementary Note 6: Tunneling superconducting gap analysis of the T-layer**

To model the DOS ($\rho_s$) of the superconducting gap observed in the T-polymorph layer, we employed a modified version of the BCS theory that takes into account the quasiparticle lifetime, as introduced first by Dynes et al..[51]:

$$\rho_s(E) = \rho_{s,N} Re\left[\frac{|E-i\Gamma|}{\sqrt{(E-i\Gamma)^2+\Delta^2}}\right]$$

(1)

where $\rho_{s,N}$ is the DOS of the sample in the normal state, $\Gamma$ is the quasiparticle lifetime broadening due to all non-thermal sources, and $\Delta$ is the superconducting gap size. For the sake of simplicity, we considered $\rho_{s,N}$ as a constant or linear function of energy. To account for nodal superconductivity, we applied an approximation described in previous works[24,52], in which the superconducting order parameter is modeled as an angular dependent variable [$\Delta = \Delta_0 cos(l\Theta)$] in equation (1). Here, $l$ is the number of nodal lines in the pairing function. Therefore, assuming an energy-independent constant DOS of our STM tip, the total differential conductance (d$I$/d$V$) of the sample measured through STS is:

$$dI/dV(V) \propto \int_{-\infty}^{+\infty} dE \frac{df(\varepsilon)}{d\varepsilon}\Big|_{\varepsilon=E-eV} \rho_s(E)$$

(2)

where $f(\varepsilon)$ is the Fermi-Dirac distribution that weights for the thermal broadening. The implementation of equation (2) to fit our experimental d$I$/d$V$ spectra was performed using the open-access code provided in ref. [24].

In Fig. S6 we show several representative SC differential conductance spectra recorded on dark (Fig. S6a-c) and bright (Fig. S6d-f) contrasted SoD clusters of the T-layer (Fig. 2a in the main text). For each curve, we carried out fitting using the equation (2) of both *s*-wave and nodal ($l$ = 3) pairing symmetry for the superconducting order parameter, and with the following procedure: $\Gamma$ and $\Delta$ are set as free-fitting parameters, and the fitting range (normally $\gtrsim \Delta$) is defined so that it minimizes the obtained fitting error of $\Delta$. As seen, the nodal symmetry fits systematically better than the *s*-wave model, and it is evidenced from three perspectives:

(1) The fitted curves using a *nodal* pair wave function qualitatively capture all the features of our experimentally measured d$I$/d$V$ curves (the coherence peaks structure and the sharp V-shaped dip), while the *s*-wave fails to do so.

(2) By using the reduced χ-squared statistics ($\chi^2/\nu$) as a probe of the goodness of the fit, we noted that the nodal fits show systematically lower values of $\chi^2/\nu$ as compared to those of *s*-wave fits. This occurred in 38 out of the 40 analyzed spectra (95% of the sampling range). All the analyzed d$I$/d$V$ curves are recorded on different regions and with different tips (statistically significant curves).

(3) The quasiparticle lifetime broadening parameter required using an *s*-wave model ($\Gamma_{avg}$ ~ 0.04 meV) is substantially larger than that for a nodal model ($\Gamma_{avg}$ ~ 0.02 meV) (Fig. S6g). This is in sharp contrast with the robustness of conventional superconductors against impurities scattering as compared to nodal-like superconductors[24]. Due to the high quality of the T-type surface termination observed in our measurements, the existence of scattering caused by impurities that may justify the large broadening required to fit the results to a *s*-wave model can be ruled out. The present analysis suggests that the quasiparticle tunneling spectra recorded on the T-layer will likely have nodes in the superconducting gap function.

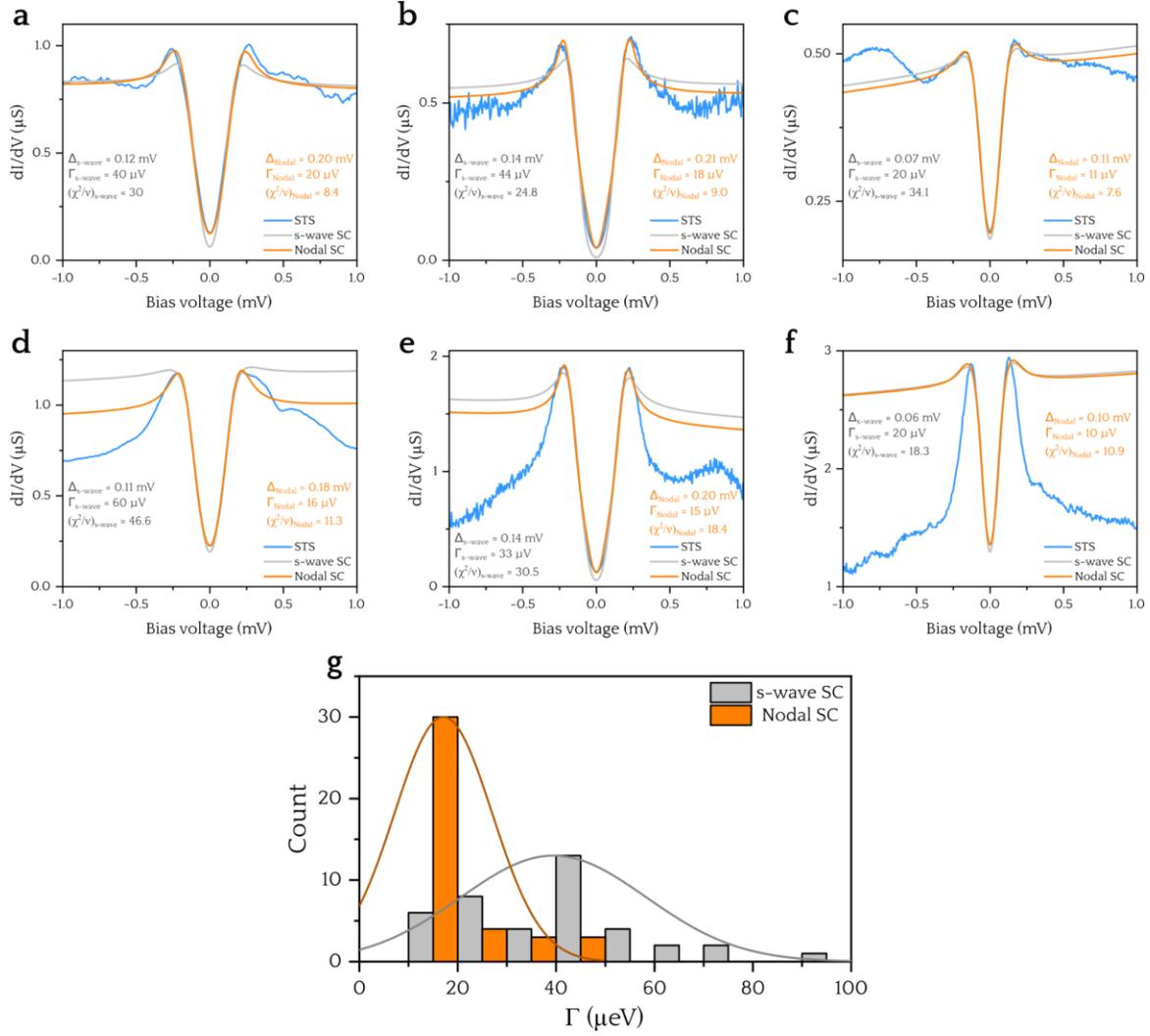

**Fig. S6 - *s*-wave and nodal nodal fits of the superconducting spectra of T-layer. a-f**, Representative d$I$/d$V$ spectra recorded on the center different SoD clusters and the corresponding fittings to the Dynes function using an order parameter with *s*-wave and nodal symmetry ($V_{a.c.}$ = 20 µV, $T$ = 0.34 K). **g**, Statistical analysis of the obtained $\Gamma$ using *s*-wave and nodal symmetries. We employed an ensemble of 40 d$I$/d$V$ curves acquired in different regions and with different tips.

**Supplementary Note 7: Tunneling superconducting gap analysis of single crystal Al(111)**

To further test the validity of the fitting results extracted in supplementary note 6, we have carried out the very same fitting procedure for d$I$/d$V$ spectra acquired on bulk single crystal of Al(111), a well-established isotropic conventional $s$-wave BCS superconductor with a $T_C$ of ~ 1.2 K, similar to that of the T-layer. In the case of Al(111), our fitting results (see Fig. S7) suggest that only an isotropic $s$-wave symmetry fit our experimental spectra, both qualitatively and also quantitatively through the reduced χ-squared statistics ($\chi^2/\nu$). This has been confirmed based on an ensemble of 40, statistically significant d$I$/d$V$ acquired at different locations and with different tip apexes on Al(111). All of them consistently demonstrate that $s$-wave fits are markedly better to those using a nodal symmetry. This result corroborates the validity of the gap analysis done for the T-layer (supplementary Note 6) that supports a nodal symmetry.

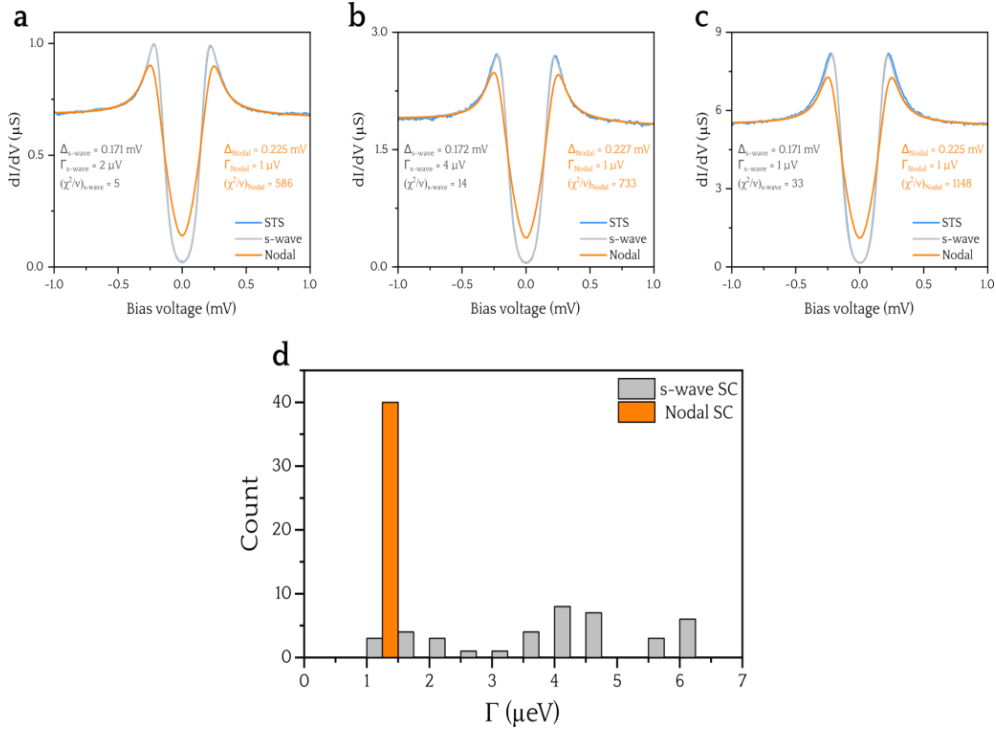

**Fig. S7 - Comparison between *s*-wave and nodal fits of the superconducting spectra of Al(111). a-c**, Representative d$I$/d$V$ spectra recorded at different location of atomically clean terraces of bulk single crystal of Al(111) and the corresponding fittings to the Dynes function using an order parameter with *s*-wave and nodal symmetry ($V_{a.c.}$ = 5 µV, $T$ = 0.34 K). **d**, Statistical analysis of the obtained Γ using *s*-wave and nodal symmetries. We employed an ensemble of 40 d$I$/d$V$ curves acquired in different regions and with different tips.

**Supplementary Note 8: Simulated Andreev reflection spectra within the framework of the Blonder-Tinkham-Klapwijk (BTK) model**

The normalized dimensionless conductance, $\sigma_{BTK}$, that accounts for the Andreev reflection spectra across the interface formed between an isotropic *s*-wave superconductor and a normal metal along the *c*-axis, is given by the 1D-BTK formula[21,23]:

$$\sigma_{BTK}(E) = \frac{1+\tau_N|\gamma(E)|^2+(\tau_N-1)|\gamma(E)^2|^2}{|1+(\tau_N-1)\gamma(E)^2|^2} \tag{3}$$

with $\gamma(E) = \frac{(E+i\Gamma)-\sqrt{(E+i\Gamma)^2-\Delta^2}}{\Delta}$ and $\tau_N = \frac{1}{1+Z^2}$. $\Gamma$ is the quasiparticles lifetime broadening, $\Delta$ is the effective pair potential (order parameter), and $Z$ is the barrier transparency parameter, which describes the goodness of the junction contact ($Z = 0$ indicates perfect contact).

To capture superconductors with sign-changing order parameters with nodes, we model the pair potential as a function of the azimuthal angle $\theta$ in equation (3)[24]: $\Delta = \Delta_0 \cos(3\theta)$ (in the case of a nodal $l = 3$ symmetry), similar to what we did with the Dynes function to describe a nodal superconductor (supplementary note 7). Then, the total conductance is obtained by integration over the whole azimuthal angle range (0 to $2\pi$):

In Fig. S8 we show simulated conductance curves using equation (3) for an isotropic *s*-wave and a nodal superconductor at various barriers strength $Z$. The Andreev reflection spectra of a *s*-wave superconductor is characterized by a steady population of states inside the SC gap until reaching a flatted-top spectrum at $Z = 0$. Meanwhile, the main characteristic for a nodal superconductor is the emergence of a peak-like structure upon reaching the perfect barrier transparency regime at $Z = 0$. We also note that a nodal *s*-wave symmetry of the order parameter leads to similar in-gap features[20].

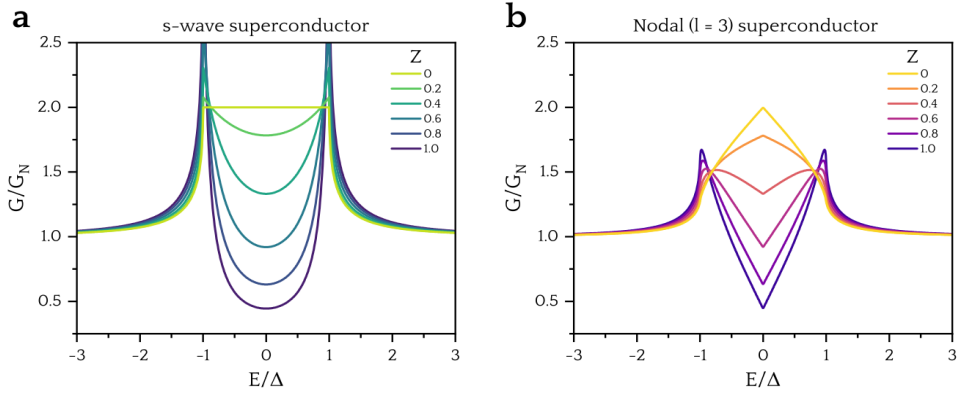

**Fig. S8 - Simulated Andreev reflection spectra in the BTK model.** Simulated normalized transport conductance spectra along the *c*-axis (orthogonal to the sample surface) in absence of broadening ($Z = 0$) between a normal tip and a *s*-wave (**a**) or nodal (**b**) superconductor for various barrier strength *Z*.

**Supplementary Note 9: Robustness of the zero-energy conductance decay in the H/T step edge**

In Fig. 3 of the main text we showed the exponential decay of the intensity of zero-bias conductance close to a step edge of T/H. Here, we prove that this decay persists also in a step edge of H/T (i.e., an inverted step edge configuration). The results are plotted in Fig. S9, indicating the robustness of this decaying behavior.

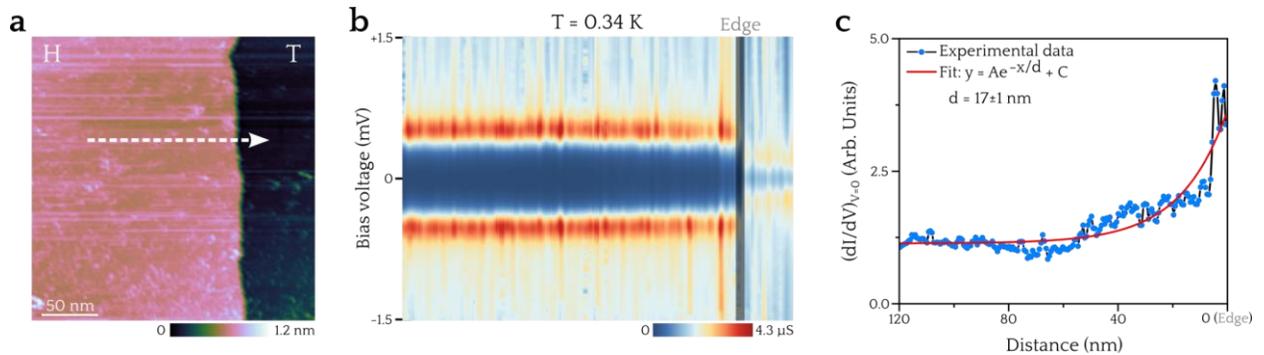

**Fig. S9 – Zero-energy conductance decay along the H/T step edges. a**, STM image displaying two consecutive H- and T-layers ($V_s = 0.1$ V, $I = 20$ pA). **b**, High-resolution tunneling d*I*/d*V*(*x*,*V*) taken along the dashed white line in **a**. **c**, Zero-energy differential conductance profile, d*I*/d*V*(*x*,0), extracted from **b**. The red line represents the fit to a first-order exponential decay. The data shown here were measured at $T = 0.34$ K.

**Supplementary Note 10: Zero-energy conductance behavior in the normal state**

The exponential decay of the zero-energy intensity along the H-type edges is not observed above $T_C$. Figure S10 shows a $dI/dV(x,V)$ profile along a H/T step edge at 4.2 K, well above $T_C$. The $dI/dV$ signal at zero energy shows a steady constant value when approaching the step edge, with a sharp increase when reaching the T-type surfaces. The present result rules out any artifact caused by the geometrical interface of the step edge, as well as any possible intrinsic edge state unrelated to superconductivity. Here we note that both polymorphs exhibit even in the normal state a shallow dip of $\simeq 2$ mV in width around $E_F$. This dip feature is tip-dependent but commonly observed in tunneling spectra around $E_F$ on TMD metals[53–56] and, while its origin is not entirely clear to date, it has been ascribed to Coulomb blockade phenomena in the tunneling junction[57] or phonon-assisted inelastic tunneling[58].

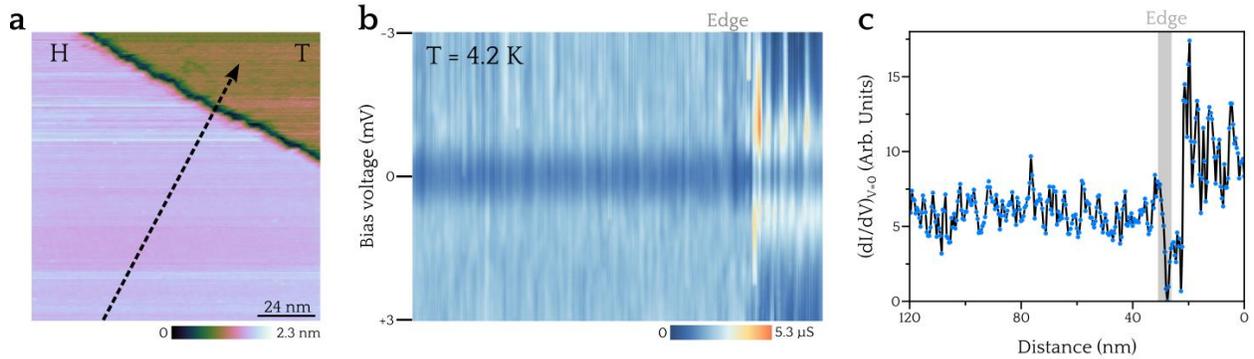

**Fig. S10 - Absence of the zero-energy conductance decay in the normal state. a,** STM image showing a H/T atomic step region ($V_s = 0.1$ V, $I = 60$ pA, $T = 4.2$ K). **b,** $dI/dV(r, V)$ profile recorded along the dashed black arrow in **a**. **c,** $dI/dV(r, V = 0)$ intensity profile extracted from **b**.

**Supplementary Note 11: Bulk superconductivity characterization of 4Hb-TaSSe crystals**

To study the bulk superconductivity in 4Hb-TaSeS, magnetization measurement was performed in both zero field cooled (ZFC) and field cooled (FC) modes at $H = 1$ mT (see Fig. S11a). The observed diamagnetic signal indicates transition into the superconducting state at onset temperature of $T_C \sim 2.1$ K with the shielding volume fraction near 25%. To determine the upper critical field parameters of 4Hb-TaSSe, field and temperature dependent magnetization measurements (see Fig. S11b,c) were conducted in two orientations: $H^\perp$ (out-of-plane field) and $H^{\|}$ (in-plane field). We roughly estimate values at $T = 0$ K from GL fit to be $H_{C2}^{\perp} \sim 2.7$ T and $H_{C2}^{\|} \sim 11.7$ T.

The estimated upper critical field value along in-plane direction, $H_{C2}^{\|}$, greatly exceeds the Pauli limit by nearly three times. The Pauli limit can be calculated as $H_p = 1.86 T_C$ giving rise to a value of $H_p = 3.98$ T for 4Hb-TaSSe.

To confirm the superconductivity in 4Hb-TaSSe, zero field resistivity and specific heat measurement were performed. The observed temperature dependent resistivity drops to zero at $T_C \sim 2.4$ K (see Fig. S11d), confirming the transition into the superconducting state. From the specific heat discontinuity at zero magnetic field (see Fig. S11e), we determine a $T_C \sim 2.15$ K, which is consistent with the magnetization and resistivity data. By fitting the normal state specific heat data using $C/T = \gamma_n + \beta T^2$, we estimated the Sommerfeld coefficient $\gamma_n$ and the Debye constant $\beta$ to be $\sim 4.8$ mJ/ mol·K$^2$ and $\sim 0.84$ mJ/mol·K$^4$, respectively.

The obtained values for bulk $T_C$ and $H_{C2}$ are in good agreement with the one inferred from local STS data, as the one shown in Fig. 4c,d in the main text. More precisely, the bulk $T_C$ matches closely to the one extrapolated for the H-surface termination, while the $H_{C2}^{\perp}$ is close to the one of the T-layer (without considering the resolution limitation of the temperature in our transport data).

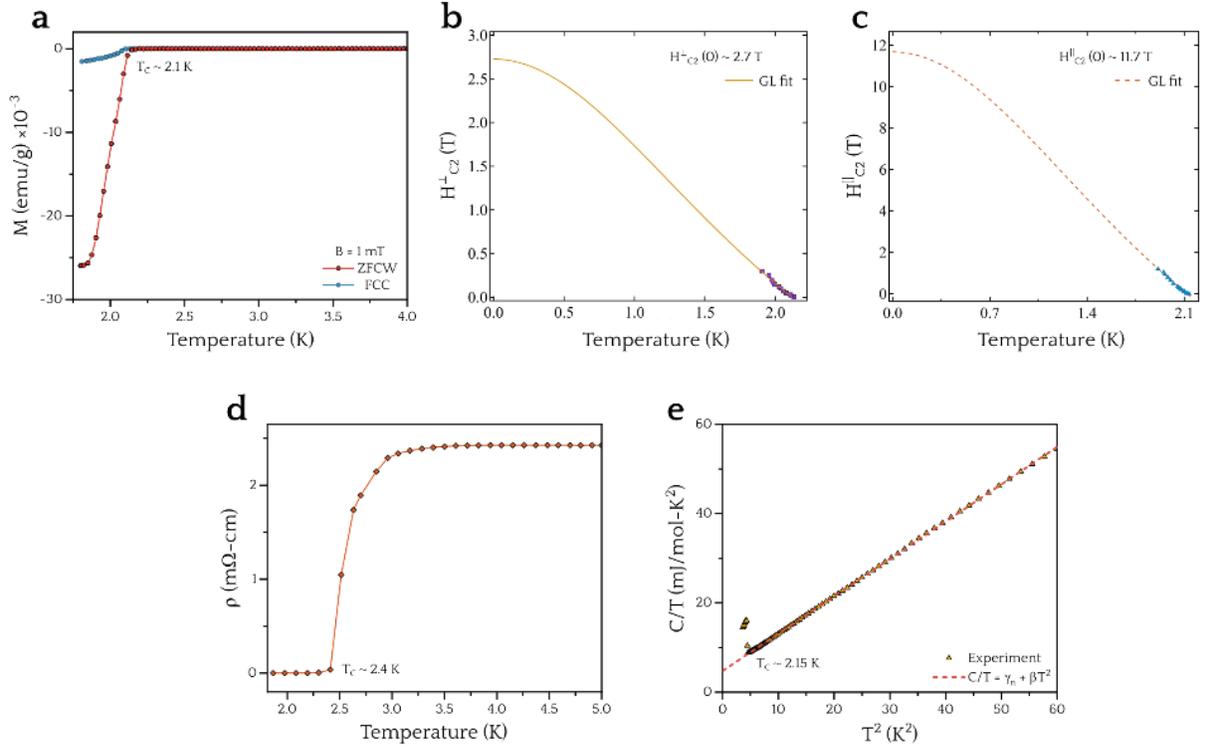

**Fig. S11 - Bulk superconductivity characterization of 4Hb-TaSSe crystals. a**, Superconducting transition temperature determined through magnetization measurement. **b,c**, Corresponding out-of-plane and in-plane upper critical field values as function of temperature extrapolated from magnetization measurements. **d**, Resistivity *vs T* experiments. **e**, Specific heat jump observed at zero field.

**Supplementary Note 12: Influence of the magnetic field direction on the SC ground state of T- and H-layer**

Previous results on heterostructures of T/H layers of Ta-based dichalcogenides[36,43] have shown the emergence of physics related to magnetism that is localized within the SoD clusters of the T-layer. Here, in order to provide insight into the possibility of the existence of magnetism in the T-layer that may influence the behavior of the SC ground state in this layer, and on the H-layer, we investigated the impact of the out-of-plane magnetic field polarity (direction ± Z) on the experimentally acquired SC d*I*/d*V* spectra. A comparative result between two representative dI/dV curves recorded at both magnetic field polarity on the T- and H-surface termination are depicted in Fig. S12a,c, respectively. From these graphs, one may already infer that the magnetic field orientation does not affect the overall SC spectra shape of the T- and H-layer. Still, another possibility is that the magnetic field polarity does affect the upper critical field values $H_{C2}$. However, we ruled out this scenario from the superconducting gap *versus* $H_\perp$ plots shown in Fig. S12b,d. From our experimental results, we estimate an upper critical magnetic field of $H_{C2}^{+Z} = 3.2 \pm 0.3$ T and $H_{C2}^{-Z} = 2.9 \pm 0.2$ T for the T layer, and $H_{C2}^{+Z} = 0.73 \pm 0.1$ T and $H_{C2}^{-Z} = 0.70 \pm 0.1$ T for the H layer, respectively. This result indicates that the magnetic field polarity has a residual impact on the SC ground state of the T- and H-surface termination.

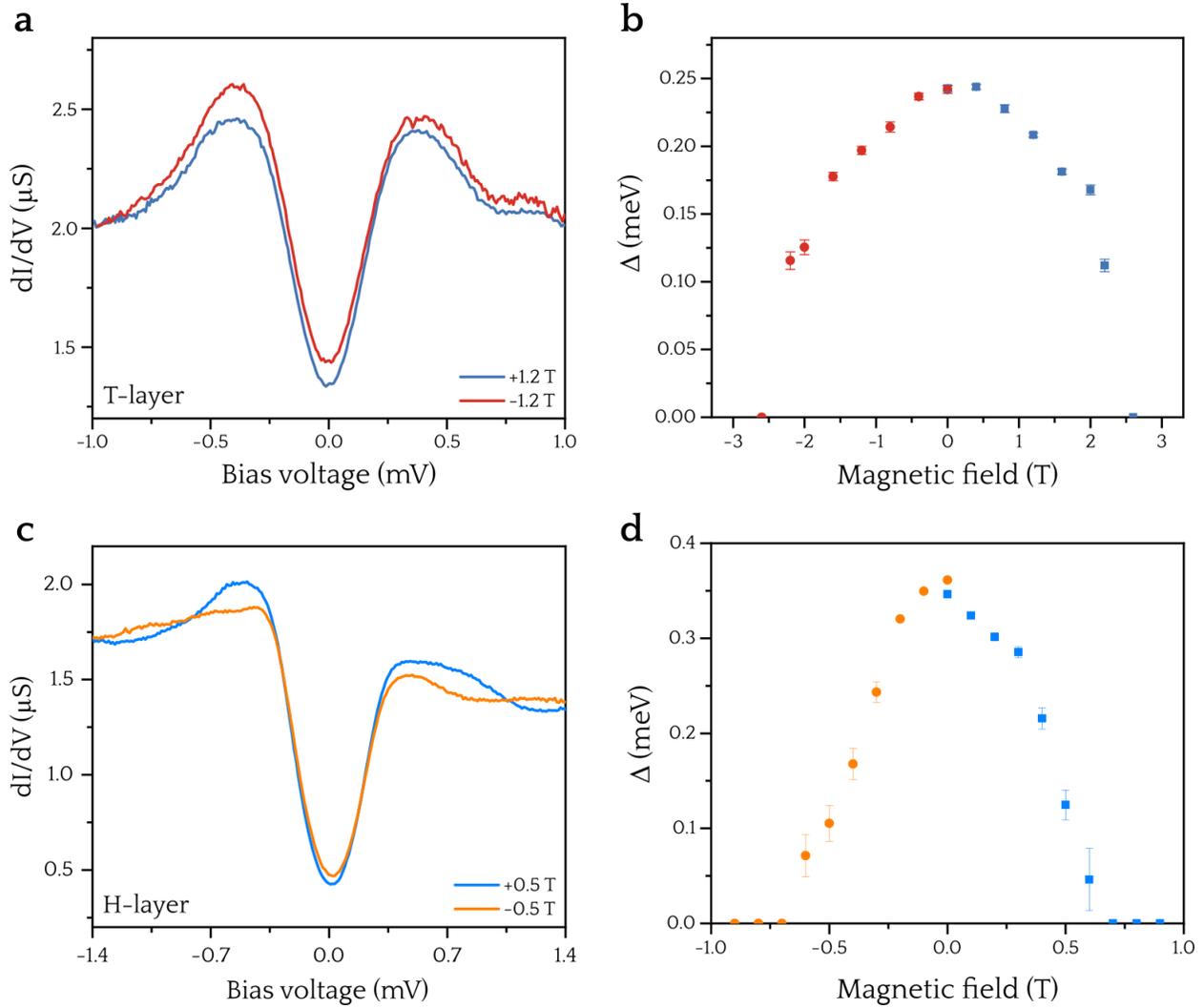

**Fig. S12 - Influence of $H_\perp$ polarity on the superconductivity of T- and H-surface termination. a,c**, High-resolution differential conductance curves recorded at both magnetic field polarity in the T- and H-layer, respectively. **b,d**, Corresponding evolution of the superconducting gap as a function of magnetic field on the T- and H-layer. Acquisition parameters: **a,b**, $V_{a.c.} = 30$ μV, $T = 0.40$ K. **c,d**, $V_{a.c.} = 20$ μV, $T = 0.40$ K.

**Supplementary Note 13: STS interpretation of dI/dV spectra on T layer with DFT calculations**

The experimentally measured electronic structure of the T layer at the center of the SoD is compared in Fig. S13 with our *ab-initio* calculated band structure of 4Hb-TaSSe in the √13×√13 CDW state (see Methods), projected to the $d_{z^2}$ orbital of the SoD Ta atom. The experiment displays two broad peaks labeled as $C_1$ and $V_1$. These correspond to the $d_{z^2}$ flat band state which remains essentially unhybridized at $k_z = 0$ ($C_1$), and to an occupied state below the flat band with strong $d_{z^2}$ weight at the $\Gamma$ point ($V_1$). Lastly, the sharper zero-bias peak at some of the SoD center locations likely originates from an inhomogeneous population of the SoD states in the surface T layer combined with correlation effects, neither of which can be captured by our *ab initio* calculations.

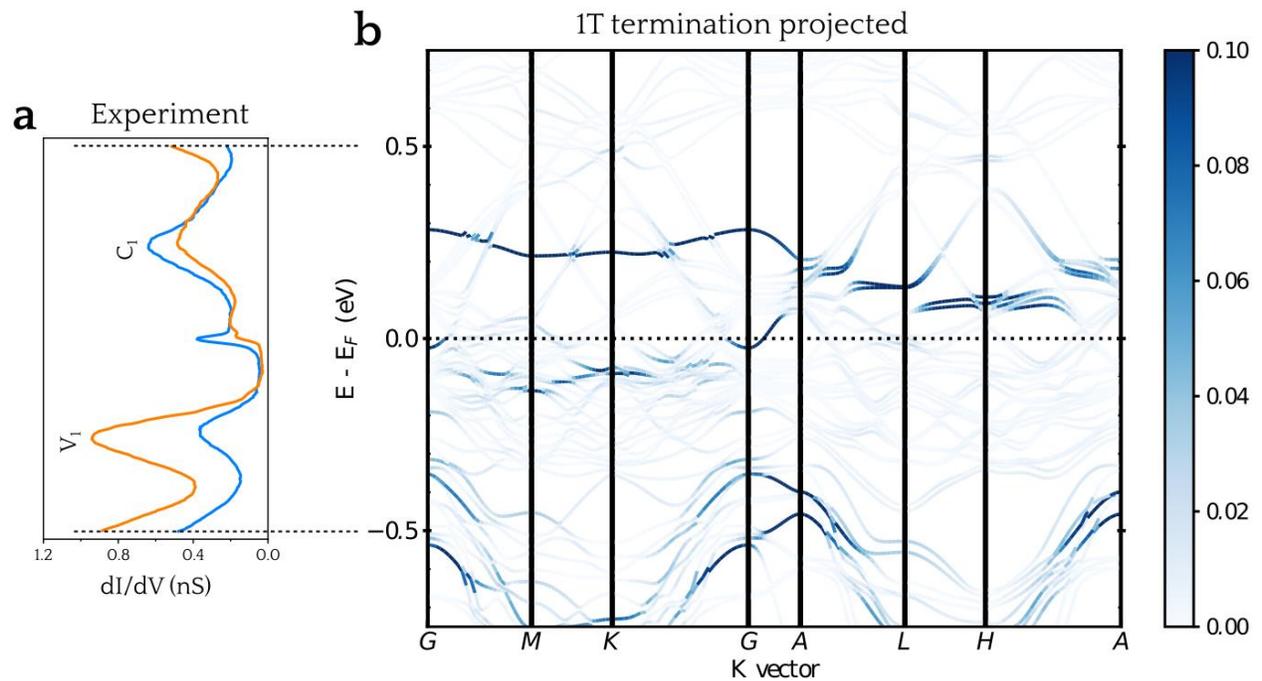

**Figure S13 - Comparison between experimental STS and theoretical band structure of T-layer. a**, Experimentally recorded large-window STS spectra on the T-layer (same as in Fig. 1f.) Acquisition parameters: $T = 4.2$ K. **b**, Band structure of 4Hb-TaSSe projected to the $d_{z^2}$ orbital of the SoD Ta atom in the T-layer, calculated within the $(√13×√13)a_0$-13.9° CDW supercell from *ab initio* methods.

## Supplementary Note 14: Spectroscopic features of the T-layer for finite interlayer hybridization

In our effective model, the finite interlayer hybridization opens a gap in the DOS projected on the T-layer, even in the absence of attraction in that layer. Fig. S14 shows the layer projected d$I$/d$V$ of the effective model for an attractive onsite interaction only for the H layer ($V_H > 0, V_T = 0$). As seen, besides the multigap feature in the H-layer, there is an induced gap in the T-layer which formally survives up to temperatures at which $\Delta_H \to 0$. The dI/dV curves are calculated with thermal broadening (a convolution of the DOS and the derivative of the Fermi function) at each temperature, as in STS experiments.

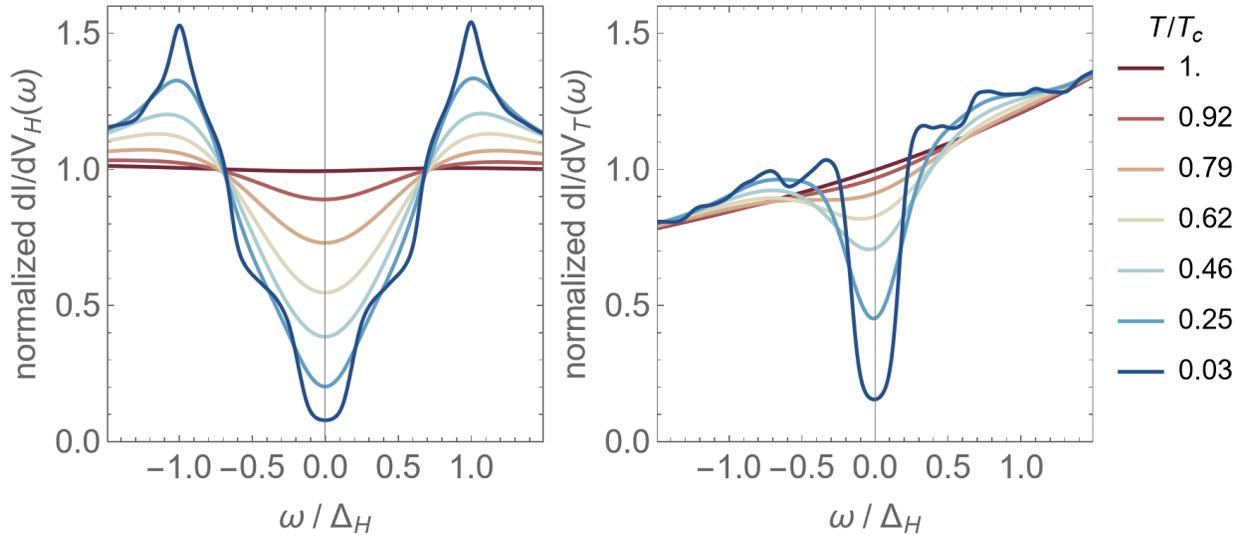

**Figure S14 - Induced gap in the T layer projected DOS due to the H-T interlayer hybridization.** Calculated dI/dV curves in the minimal model for $V_H > 0, V_T = 0$ (i.e. $\Delta_H > 0, \Delta_T = 0$ for all $T < T_C$) with hybridization $t_z = 10\Delta_H(T = 0)$ at selected temperatures up to $T_C$ (specified in legend), projected on the H-layer (left) and the T-layer (right).